\documentclass[12pt]{article}
\usepackage{graphicx}
\usepackage{dcolumn}
\usepackage{bm}
\usepackage{color}
\begin{document}
\bibliographystyle{unsrt}
\begin{center}  {\Large \bf Mach-Zehnder interferometer with absorbing Fabry-P\'{e}rot cavities}
\\~\\ {\normalsize  Victor-Otto de Haan}  \\~\\
{\it BonPhysics B.V., Laan van Heemstede 38, 3297 AJ Puttershoek, The Netherlands}~ \\
{\it \today} \end{center}
\section*{Abstract}
A Mach-Zehnder interferometer with absorbing Fabry-P\'{e}rot cavities is used to measure the optical phase differences upon rotation around a vertical axis. The details of the expected experimental results are described based on the idea that a conducting medium can introduce phase differences due to changes in material conductivity that are expected to occur when a medium moves with respect to a reference frame. Experimental details and results are discussed.

\section{Introduction}
In Einstein's theory of special relativity~\cite{Einstein1905} it is assumed that light propagates in empty space with a constant velocity independent of the velocity of the observer. Many experiments support this assumption. However, it is well know that most of these experiments are also supportive of the assumption that light propagates through a medium under conditions of time dilatation and Lorentz contraction. That light, or in general electro-magnetic waves, can be interpreted as waves propagating through a medium is the basis of the success of the Maxwell equations. Lorentz~\cite{LorentzElectronBook} was able to show that Maxwell equations could lead to Lorentz contraction and time dilatation for systems moving with respect to the medium. When the medium is referred to as {\it ether}, this is known as the Lorentz ether theory. Until recently is was thought that Einstein's theory of special relativity and Lorentz ether theory were experimentally indistinguishable. The preferred reference frame or ether frame could never be detected and hence ignored. However, it has become clear that in principle experiments are possible to distinguish between them~\cite{Kholmetskii2014},~\cite{deHaanBoek}.

One type of such experiments is based on the occurrence of Wigner rotation or Thomas precession due to the fact that two successive non-co-linear Lorentz transformations can only be described by means of a Lorentz transformation of the composite velocity and a rotation over the Wigner rotation angle~\cite{Ungar1989}. This in principle should enable the detection of the ether frame. 

Another type is based on experiments where light has some interaction with matter. As soon as a medium interacting with electro-magnetic waves exists, it is possible that this interaction enables the detection of the ether frame. A theoretical possibility has recently been outlined by Spavieri~\cite{Spavieri2008} and Consoli~\cite{Consoli2013} although the first one has been disproved by experiment~\cite{deHaan2010}. Another possibility is the occurrence of superluminal signal transport, as assumed to be possible in quantum mechanics due to its non-local character as discussed for instance by Einstein~\cite{Einstein1935} for the Einstein-Podolsky-Rosen thought experiment. 

The absorption process of light inside a conductor is poorly understood. It is based on Ohm's law relating the free current in a conductor with the electric field strength, interpreted as a resistance against the vibrations of the electrons. In classical theory this resistance can be interpreted as due to scattering of electrons at certain time intervals, so that the electron momentum and energy is transferred to the molecules, ions or atoms constituting the medium.
It can also be interpreted as a {\rm frictional} force due to interaction of the electron with the atom to which it is bounded. In current theories quantum mechanics is involved to find models for the resistivity. However, none of these models are proven to be explicitly Lorentz co-variant and hence conductivity is a material property that {\it could} depend on the velocity of the material with respect to the ether.

According to some authors~\cite{Nimtz2002},~\cite{Nimtz1998},~\cite{Missevitch2011},~\cite{Longhi2003} the interaction between light and matter results in superluminal signal transport. Longhi~\cite{Longhi2003} shows that light transmitted through two fiber Bragg gratings at some distance apart experiences a phase shift independent of this distance. Olkhovsky~\cite{Olkhovsky2002} argues that this also holds for resonant tunneling through a double barrier. Each barrier could be build from a fiber Bragg grating as was done by Longhi. In such a case the wavelength must be carefully tweaked so that the transmission is small (but not too small). However, any double barrier system (for light or quantum-mechanics) can experience such a phase behavior as long as the total transmission is small. In principle there is no reason why the barriers should consist of fiber Bragg gratings. A simple metal layer on a glass substrate (appropriately chosen thickness) gives similar results as has been shown by for instance Monzon~\cite{Monzon1996}, Giust~\cite{Giust2000} and de Haan~\cite{deHaan2011}. 

The above arguments can be combined and resulted in the experiment described below.

\section{Refractive index of a conducting material}
In a simple model of a layer made of a conductor with a conductivity $\sigma$, the plane wave solution of Maxwell's equations can be developed from the Lorentz-Drude model~\cite{Lorentz1895}, where the relation between the polarization of the material, $\vec{P}(\vec{r},t)$ is proportional to the distance of each electron $\vec{x}_i$ to its position when the electric field $\vec{E}(\vec{r},t)$ would be zero. This gives
\[
\vec{P}(\vec{r},t) = n_e q \vec{x}(\vec{r},t)
\]
where $n_e$ is the electron density of the material, $q$ the electric charge and $\vec{x}(\vec{r},t)$ the average distance to the equilibrium position of all electrons. Here, it is assumed that the number of electrons in a region where $\vec{P}(\vec{r},t)$ is (almost) constant is large enough to enable a statistical averaging with small enough standard deviation. The equation of motion of each electron is assumed to be (ignoring the magnetic field and the magnetic interactions)
\[
m_e\frac{d^2\vec{x}(\vec{r},t)}{dt^2} + k_e \vec{x}(\vec{r},t) + f\frac{d\vec{x}(\vec{r},t)}{dt} = q\left(\vec{E}(\vec{r},t) +  \frac{a}{\epsilon_o}\vec{P}(\vec{r},t)\right)
\] 
where $\epsilon_o$ is the permittivity of vacuum, $m_e$ is the electron mass and $k_e$ is an elastic constant of an elastic force driving the electron back to its equilibrium position. $a$ is due to the interaction with all other electrons and ions in the material and for a spherical cavity in a homogeneous polarized material equal to approximately 1/3. The friction coefficient $f$ can be interpreted as due to the collisions of the electrons with the atoms. If the mean time between collisions is $\tau$ then $f = m_e/\tau$.

A conduction current $\vec{J}=qn_e\vec{v}$ can be interpreted as a flow of free electrons ($a = k_e/(q^2n_e)$) moving with constant uniform speed $\vec{v}$ so that the acceleration is zero, hence the above reduces to $f\vec{v} = q\vec{E}$. According to Ohm's law $\vec{J}=\sigma\vec{E}$ so that $f = q^2n_e/\sigma$. Note that this interpretation does not hold for conditions where the electrons are not free, as then the friction coefficient might be small although the global conduction is negligible. 

In the harmonic case the above equation reduces to
\[
\left( - \omega^2/(\omega_p^2\epsilon_o) - a'/\epsilon_o + i\omega/\sigma \right) \vec{P}(\vec{r},t) = \vec{E}(\vec{r},\omega)
\] 
where $\omega_p$ is called the {\it plasma frequency} equal to $q\sqrt{n_e/(m_e\epsilon_o)}$ and $a'= a - k_e/(q^2n_e)$. Under consideration that $\vec{D}(\vec{r},\omega) = \epsilon_o\vec{E}(\vec{r},\omega)+ \vec{P}(\vec{r},\omega) $ and the constitutive relation between the electric field and the electric displacement $\vec{D}(\vec{r},\omega) = \epsilon_r(\omega) \epsilon_o \vec{E}(\vec{r},\omega)$ gives the relative permittivity of the material as
\[
\epsilon_r(\omega) = 1 + \frac{1}{-\omega^2/\omega_p^2  - a' + i \kappa \omega/\omega_p} 
\] 
where $\kappa = \omega_p \epsilon_o/\sigma = 1/(\tau \omega_p)$. 

The refractive index of the material is given by
\[
n(\omega) = \sqrt{\mu_r\epsilon_r(\omega)} = n_r + i n_i
\]
where $\mu_r$ is the relative permeability of the material, $n_r$ the real part of the refractive index and $n_i$ its imaginary part. As before, the magnetic interactions are ignored so that $\mu_r=1$. For a good conductor both $a'<<1$ and $\kappa <1$ so that the refractive index can be approximated by 
\[
n_r = \frac{\kappa}{2\omega_r^2\sqrt{1-\omega_r^2}}\left(1-\frac{\kappa^2+3a'}{2\omega_r^2}\right) \]\[
n_i = \frac{\sqrt{1-\omega_r^2}}{\omega_r}\left(1-\frac{\kappa^2+a'}{2\omega_r^2}\right)
\]
for $\omega_r< 1$ and by
\[
n_r = \frac{\sqrt{\omega_r^2-1}}{\omega_r} \]\[
n_i = \frac{\kappa}{2\omega_r^2\sqrt{\omega_r^2-1}}
\]
for $\omega_r>1$. Note that for $\omega_r < 1$ the real part of the refractive index is proportional to $\kappa$, hence inversely proportional to the conductivity. 

\begin{figure}
\begin{picture}(400,240)
\put(10,00){\scalebox{0.5}{\includegraphics{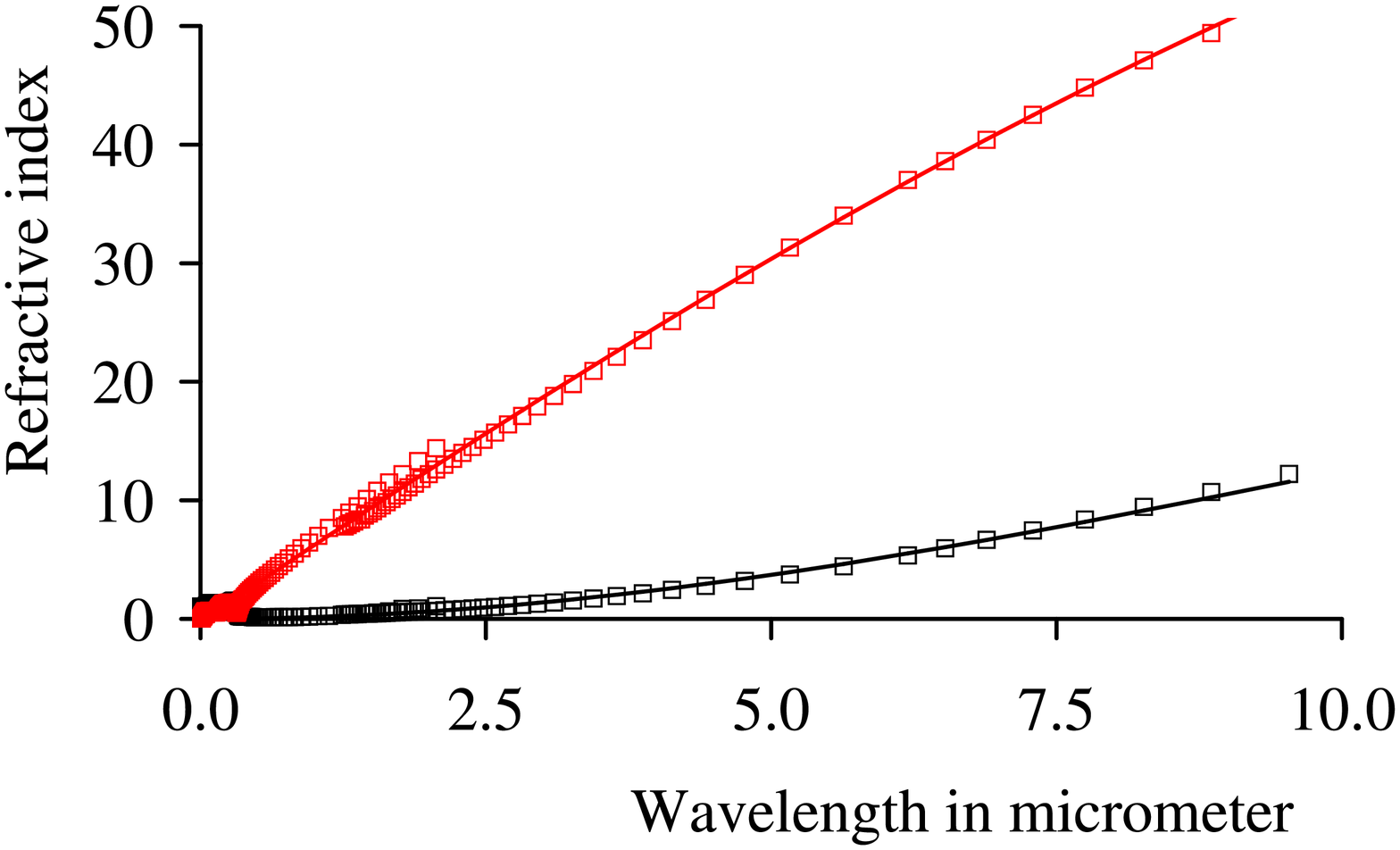}}}
\put(60,132){\scalebox{0.1825}{\includegraphics{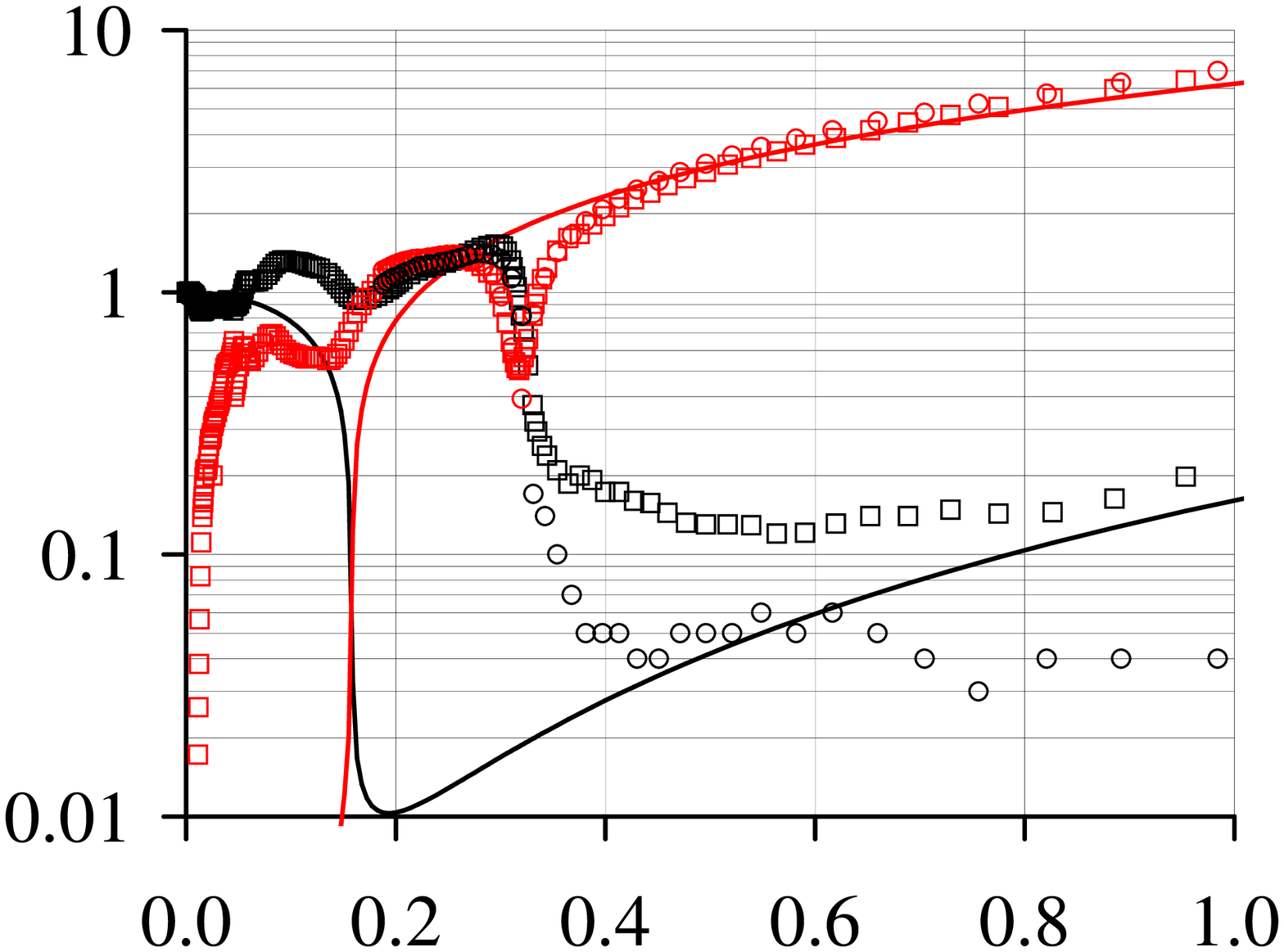}}}
\end{picture}
\caption{\label{figRefracAg} Real (black) and imaginary (red) part of the refractive index of Silver as function of wavelength as reported by~\cite{RefAg1} (circles) and \cite{RefAg2} (squares) compared to the simple Lorentz-Drude model as described in the text. The inset is the same with a logarithmic scale for the refractive index to make clear that for short wavelength the model collapses.}
\end{figure}

The real and imaginary refractive index for bulk Silver~\cite{RefAg1},\cite{RefAg2} as function of wavelength (in vacuum) is shown in figure~\ref{figRefracAg}. For longer wavelength the above model of the refractive index is fitted to these data and shown as the solid lines. The plasma frequency $\omega_p=1.2\times 10^{16}$~1/s, or 7.8~eV corresponding to a wavelength of 0.15~$\mu$m. This is similar to the value reported by Raki\'c~\cite{Rakic1998} (9.0~eV), but he used a far more extensive model. Further $a'=3\times 10^{-5}$ and the fitted conductivity $\sigma=1.3\times 10^7$~S/m so that $\kappa=8\times 10^{-3}$. The fitted conductivity is smaller than the DC value of $6.30\times 10^7$~S/m at 20$^o$C~\cite{crc}. The bound electrons that are capable of oscillations experience a larger frictional force than the free electrons responsible for the DC conduction. For smaller wavelength the model is not correct any more. It could be improved by incorporating more oscillators (as has been done for instance by~\cite{Rakic1998}). The wavelength dependence of the refractive index for some other metals with a good conductivity (such as Copper, Gold and Aluminum) is comparable to the one of Silver. For the visible light region it can be fitted to the same model, yielding a slightly different plasma frequency. 

\section{Absorption in a conducting layer}
The general solution for a propagating plane wave in a homogeneous medium is
\[
\Psi(\vec{r},t) = \widehat{\Psi}e^{\pm\alpha(\vec{k}\cdot\vec{r})}\cos\left(\omega t \pm\vec{k}\cdot\vec{r}+\phi \right) 
\]
where $\vec{r}$ is the position vector, $t$ is time, $\Psi(\vec{r},t)$ represents any component of the electric or magnetic field, $\widehat{\Psi}$, the amplitude of the plane wave at the origin, $\phi$, the phase of the wave at the origin, $\omega$ the frequency of the wave and $\vec{k}$ the wave vector in the material. The $\pm$ indicates the direction of the plane wave, $-$ in the direction of $\vec{k}$ and $+$ against it. The dispersion relation is given by the real part of the refractive index
\[
\frac{k c}{\omega} = n_r
\]
where $c$ equals the velocity of light in vacuum. The extinction coefficient $\alpha$ is given by the ratio between the imaginary and real parts
\[
\alpha = \frac{ n_i }{ n_r }
\]
The electro-magnetic interaction of light with inhomogeneous materials can be calculated by division in regions where the material is homogeneous, solving Maxwell's equations in each region and connect the solutions at the boundaries~\cite{deHaan2011},\cite{Monzon1991}. In this way, the reflection, absorption and transmission of any optical system can be calculated and compared to experiment.

\begin{figure}
\begin{picture}(400,220)
\put(20,20){\scalebox{0.5}{\includegraphics{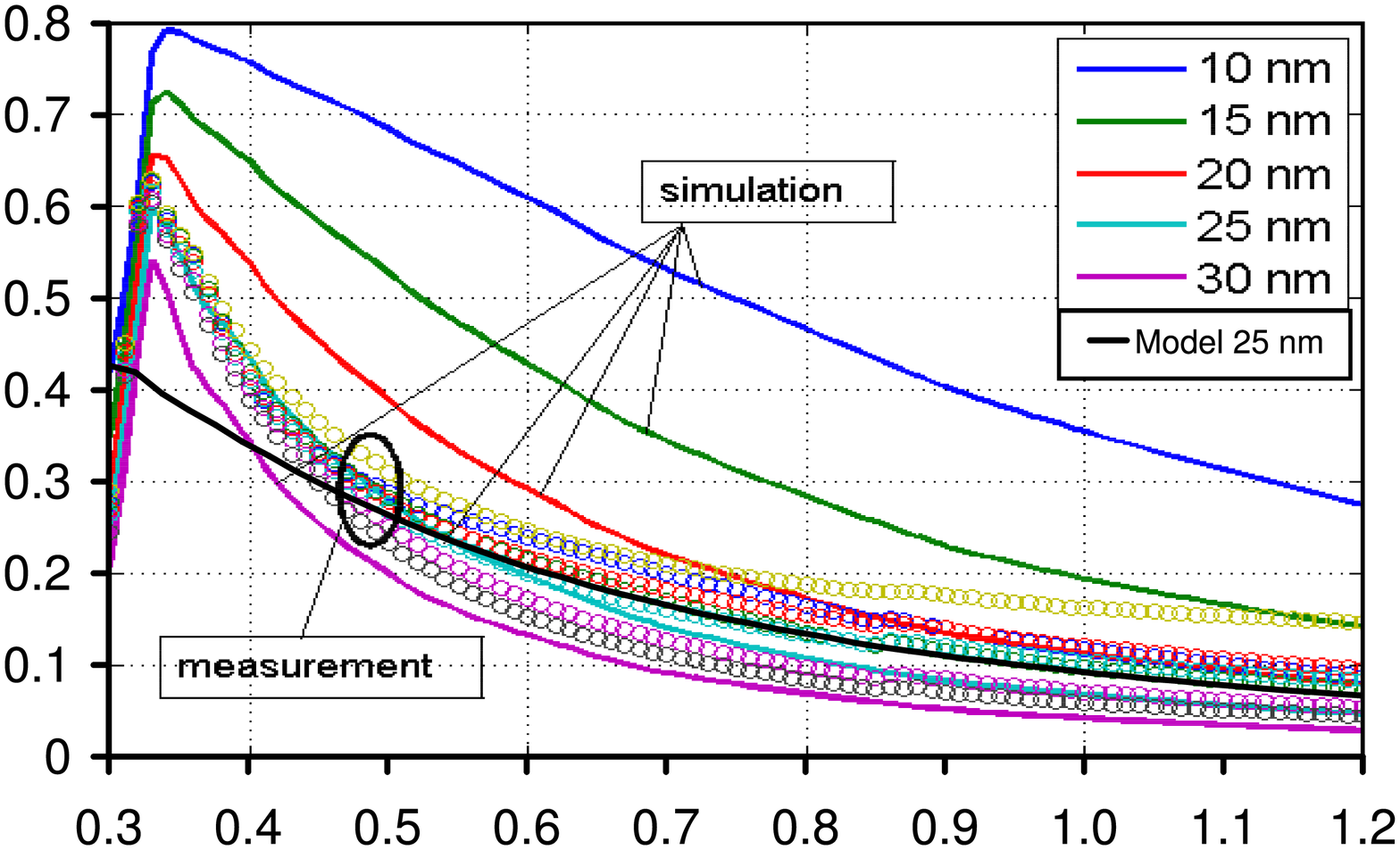}}}
\put(240,0){Wavelength in $\mu$m}
\put(0,230){Transmission}
\end{picture}
\caption{\label{figAgSamples} Wavelength dependent light-transmission measurements of several Silver layers on top of a BK7 glass substrate (circles) compared to simulations for several layer thicknesses (colored lines). The black line represents the results for a 25 nm silver layer calculated with the simple dispersion model as described in the text.}
\end{figure}

\begin{figure}
\begin{picture}(400,100)
\put(10,0){\scalebox{1.0}{\includegraphics{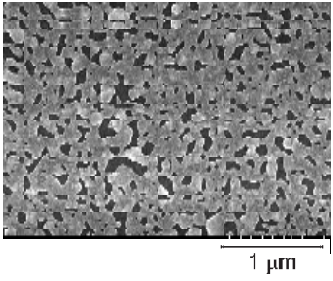}}}
\end{picture}
\caption{\label{figSEMAgLayer} Scanning electron microscope picture of a 25 nm Silver layer on top of a BK7 glass substrate. The silver layer is created by islands visible as the bright parts.}
\end{figure}

For instance, wavelength dependent measurements of the transmission of a thin silver layer on top of a 3~mm thick BK7-glass substrate are shown in figure~\ref{figAgSamples} and compared with calculations for several layer thickness using the index of refraction as given by the data points in figure~\ref{figRefracAg}.
It is clear that the transmission depends very strongly on the wavelength. Qualitatively, the transmission corresponds to that of a Silver layer with a thickness between 20 and 30~nm on top of a glass substrate of 3~mm. The black line represents the calculations for the simple dispersion model as described before and a Silver layer thickness of 25~nm. 
Again for smaller wavelength calculations differ considerably from the measurements. This is presumably because the used model refractive index differs from the actual one. 
Deviations in the refractive index can occur because the Silver layer is far from perfect. The layer tends to form small islands. A scanning electron microscope picture of one of the layers is shown in figure~\ref{figSEMAgLayer}. The islands are clearly visible as the bright parts. Here, it is assumed that the effect of the islands can be incorporated by an effective thickness and refractive index averaged over the surface of the layers. As can be seen from figure~\ref{figAgSamples} these parameters will change from sample to sample. In the following it is assumed that the refractive index of Silver is given by the Lorentz-Drude model with the fitted parameters mentioned above.  

Let $\phi$ be the phase a light beam acquires when traversing a length $L$ through the material. Then, the phase difference between phases acquired by a wave traversing a length $L$ through the material when the conductivity changes with $\Delta \sigma$ is
\[
\Delta \phi = S_\sigma\frac{\Delta \sigma}{\sigma}
\]
where
\[
S_\sigma = \sigma \frac{d\phi}{d\sigma} 
\]
is called the {\it conductivity sensitivity} of the light beam phase to a relative change in the conductivity. A similar {\it plasma frequency sensitivity} can be introduced to indicate the change of phase when the plasma frequency is changed 
\[
S_{\omega_p} = \omega_p\frac{d\phi}{d{\omega_p}} 
\]

For a length of 25 nm through Silver for a wavelength of 0.633~$\mu$m $S_\sigma = -n_rkL = -0.016$~rad and $S_{\omega_p}=0.031$. The plasma frequency sensitivity is larger than the conductivity sensitivity due to the different dependency of the real part of the refractive index on $\sigma$ and $\omega_p$. 

When passing through the Silver the beam of light is reduced in intensity, as the transmission is $T=exp(-2n_ikL)$ in this situation $T = 0.146$. By increasing the thickness the sensitivity increases but the beam intensity decreases exponential. The optimal performance depends on the minimal transmission for which the phase can still be determined with sufficient accuracy. Let us assume that this transmission is given by $T_{\rm min}$, then $kL_{\rm max}= ln(1/T_{\rm min})/(2n_i)$ and $S_{\sigma , {\rm max} } = ln(1/T_{\rm min})/(2\alpha)$. Hence, for an optimal performance the ratio between the real and imaginary part of the refractive index should be a large as possible and as such the extinction coefficient as small as possible. One should keep in mind that this rule only applies for materials for which the main part of the refraction is determined by the dissipative process of conduction. In such a case the extinction coefficient is given by
\[
\alpha = \frac{2\omega_r}{\kappa}(1-\omega_r^2) \left(1+\frac{2a'}{2\omega_r^2}\right)
\]
for $\omega_r<1$ and by
\[
\alpha = \frac{\kappa}{2\omega_r(\omega_r^2-1)}
\]
for $\omega_r>1$. For a frequency close to the plasma frequency ($\omega_r\approx 1$) the extinction coefficient changes drastically from very small values for frequencies above the plasma frequency ($\alpha=1.3\kappa$) to a maximum value at $\omega_r=1/\sqrt{3}$ ($\alpha=0.77/\kappa$). In the transition region around the plasma frequency (with a width of $\omega_p\kappa$) the extinction coefficient has a discontinuity from very small to very large values for increasing frequency. However, for frequencies above the plasma frequency the real part of the refractive index does not depend on the conduction any more and hence there is no phase change when the conduction changes. Only when the frequency is below the plasma frequency the real part of the refractive index of the material depends on the conductivity.

The transmission and sensitivities for a more complex model can be determined by numerical calculations following the method outlined in~\cite{deHaan2011}. For the sensitivity first, the phase of the transmitted wave is calculated for the model at hand, yielding $\phi_1$. Then, the conductivity of the model is increased by a small amount $\Delta \sigma$ (about 1~ppm) and the phase is calculated again, yielding $\phi_2$. Then, the sensitivity can be calculated according to $S_\sigma=(\phi_2-\phi_1)\sigma /\Delta \sigma$, yielding in this case $S_\sigma=0.062$~rad. For a light beam traversing a 25 nm Silver layer in air the transmission increases a bit $0.154$ due to the interference between forward and backward traveling beams in the Silver layer. The conductivity sensitivity, $S_\sigma$ changes sign and increases to 0.031, while the plasma frequency sensitivity in creases to a larger value of 0.81. The larger increase of the plasma frequency sensitivity with respect to the conductivity sensitivity is that due to the interference between the forward and backward light beams the imaginary part of the refractive index becomes important for the phase of the light beams. The relative influence of the plasma frequency change on the imaginary and real part is if the same order, however as the imaginary part is much larger, the absolute change in phase will also be much larger.

\begin{figure}
\begin{picture}(400,180)
\put(0,0){
\put(0,15){\scalebox{0.35}{\includegraphics{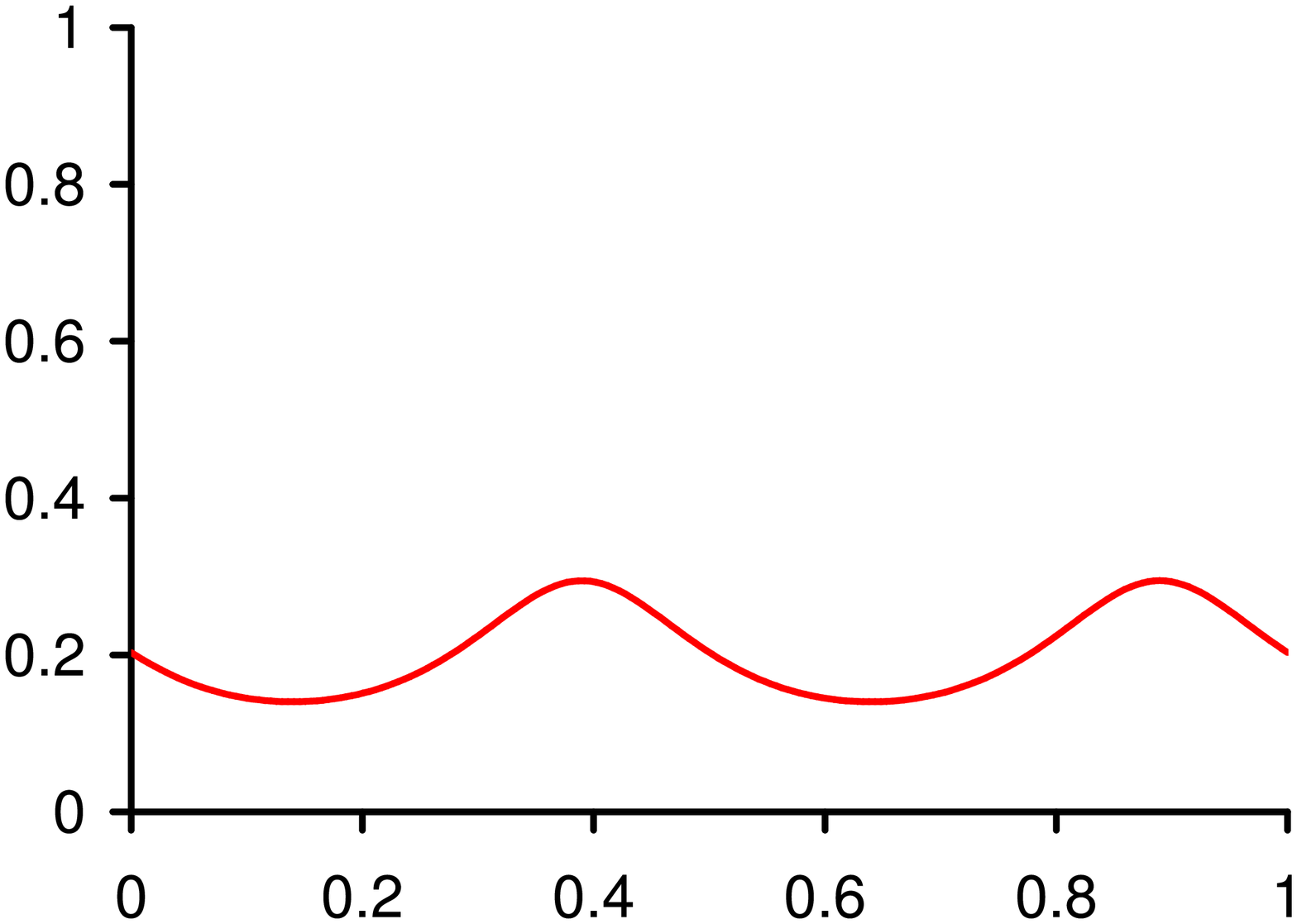}}}
\put(80,0){Relative glass thickness}
\put(0,170){Transmission}}
\put(240,0){
\put(0,15){\scalebox{0.35}{\includegraphics{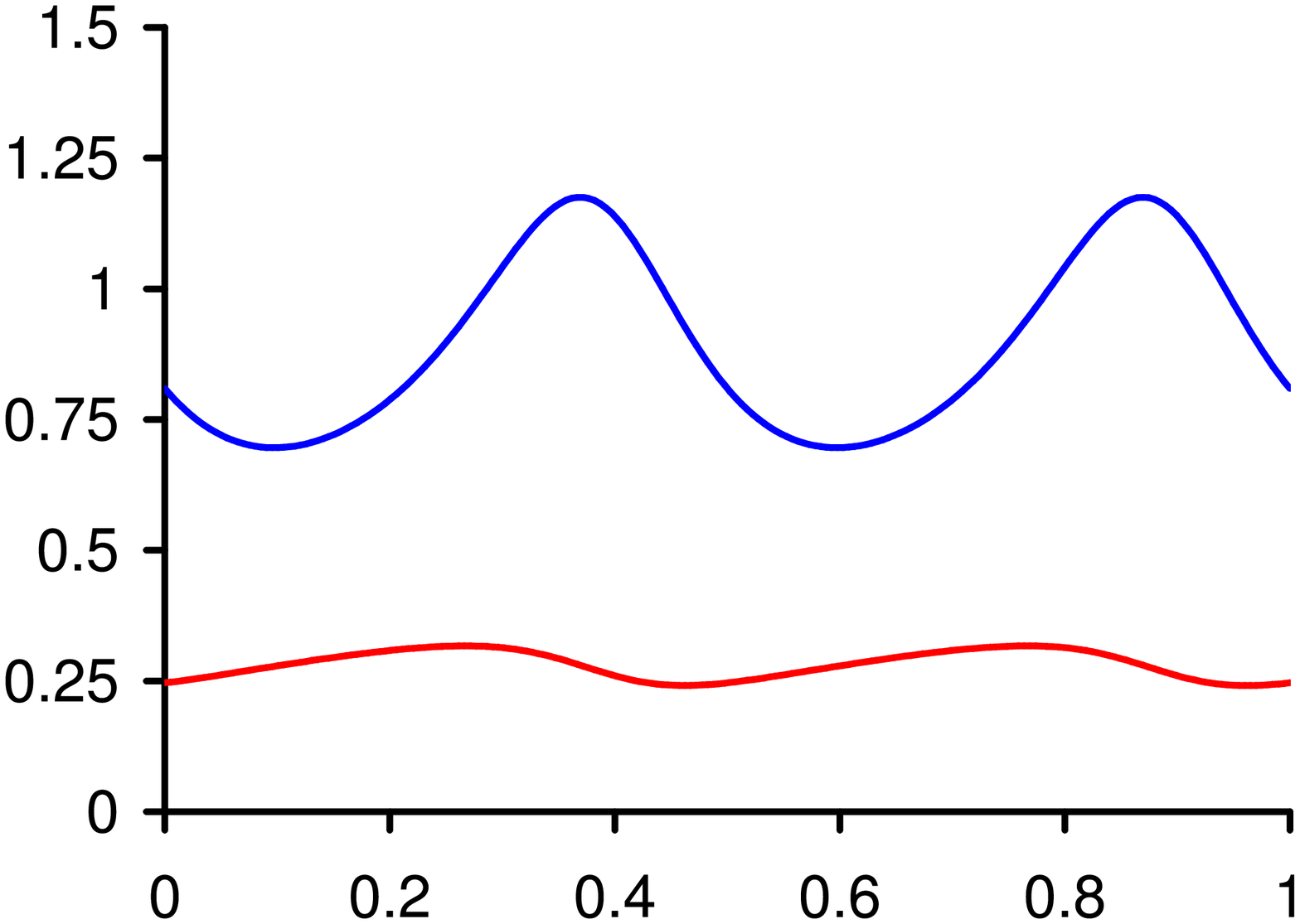}}}
\put(80,0){Relative glass thickness}
\put(0,170){Sensitivity in rad}}
\end{picture}
\caption{\label{figAgOnGlass} Left: Transmission of light passing through a Silver layer of 25 nm on a BK7 glass substrate as function of the relative glass thickness (this is the fractional part of the thickness expressed in wavelength). Right: Sensitivity due to variation of conduction (red line divided by 10) or plasma frequency (blue line) of the same model.}
\end{figure}

Such a thin layer must be deposited on a glass substrate, influencing the interference further. Then the sensitivity and the the transmission will depend on the glass thickness. For a Silver layer of 25 nm thickness on BK7 glass of a thickness varying between 3~mm and 3.004~mm (corresponding to a wavelength range in glass) the transmission and sensitivity is shown in figure~\ref{figAgOnGlass}. Depending on the glass thickness the transmission varies between 0.14 and 0.3, the conductivity sensitivity, $S_\sigma$ varies between 0.024 and 0.032 rad and the plasma frequency sensitivity between 0.7 and 1.2. For a real glass substrate the thickness will not be constant and hence the transmission and sensitivity must be averaged over the applicable thickness range.

The effect can be enhanced by putting additional layers of Silver in series in the light beam. When two of these layers are put in series they constitute an absorbing Fabry-P\'{e}rot cavity.

\section{Absorbing Fabry-P\'{e}rot cavity} \label{secFB}
\begin{figure}
\begin{picture}(450,140)
\put(100,0){\scalebox{0.6}{\includegraphics{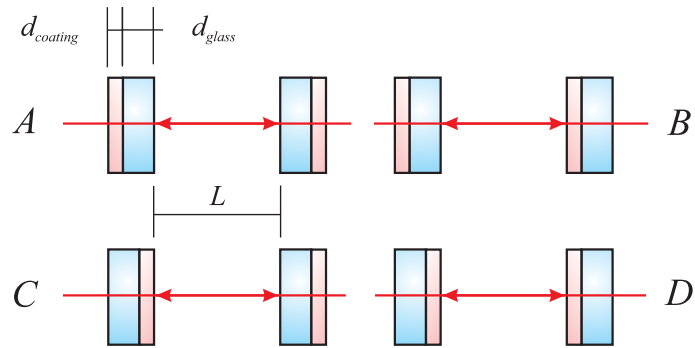}}}
\end{picture}
\caption{\label{figFBCavity} Schematic of Fabry-P\'{e}rot cavity with length, $L$. Each semi-transparent mirror consists of a thin metal layer on top of a glass substrate. The mirrors can be positioned in four different ways to form the cavity: (A) glass substrates facing each other, (B) and (C) glass substrate facing metal layer and (D) metal layers facing each other. }
\end{figure}

\begin{figure}
\begin{picture}(400,180)
\put(0,0){
\put(0,15){\scalebox{0.35}{\includegraphics{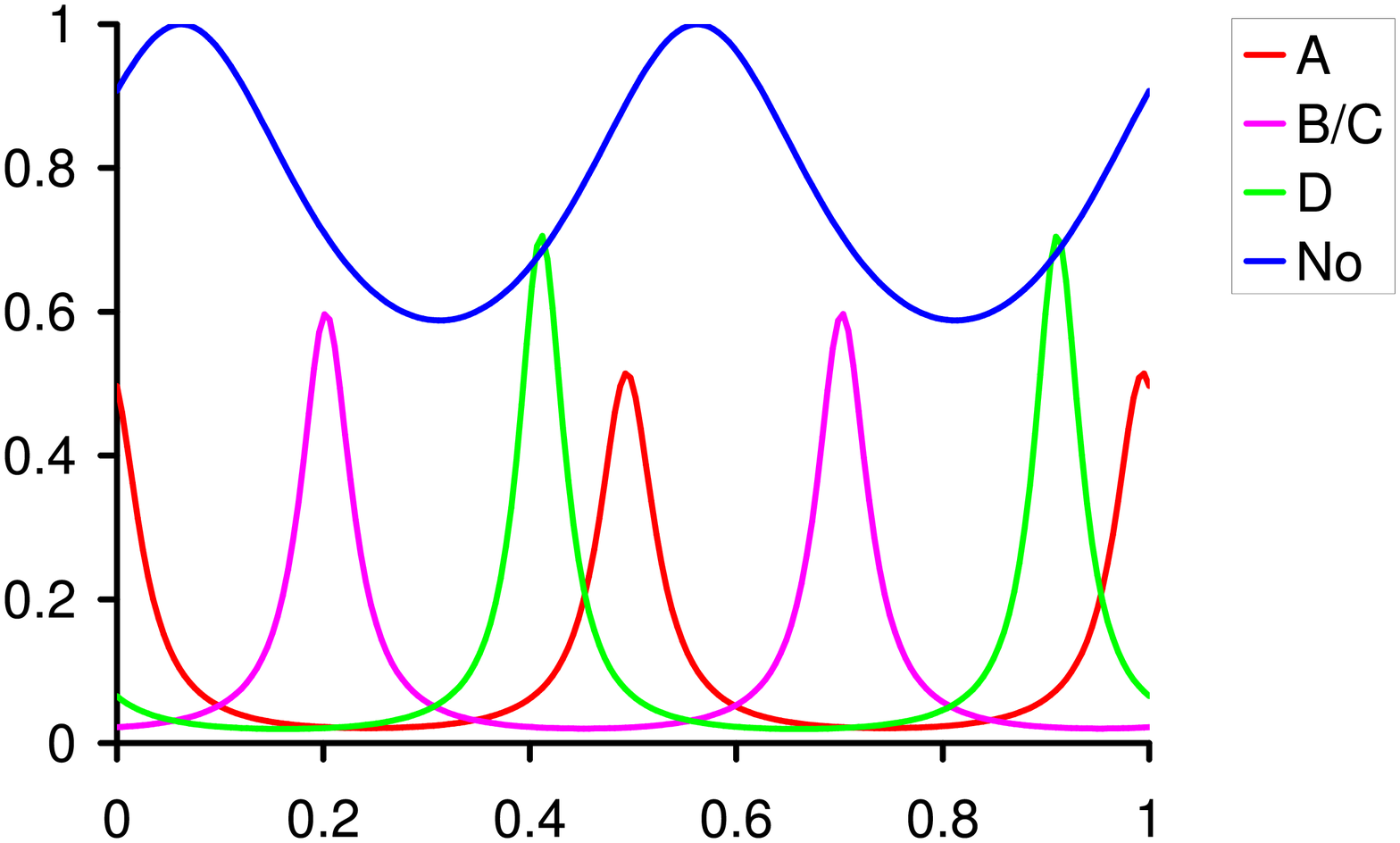}}}
\put(80,0){Relative cavity length}
\put(0,170){Transmission}}
\put(240,0){
\put(0,15){\scalebox{0.35}{\includegraphics{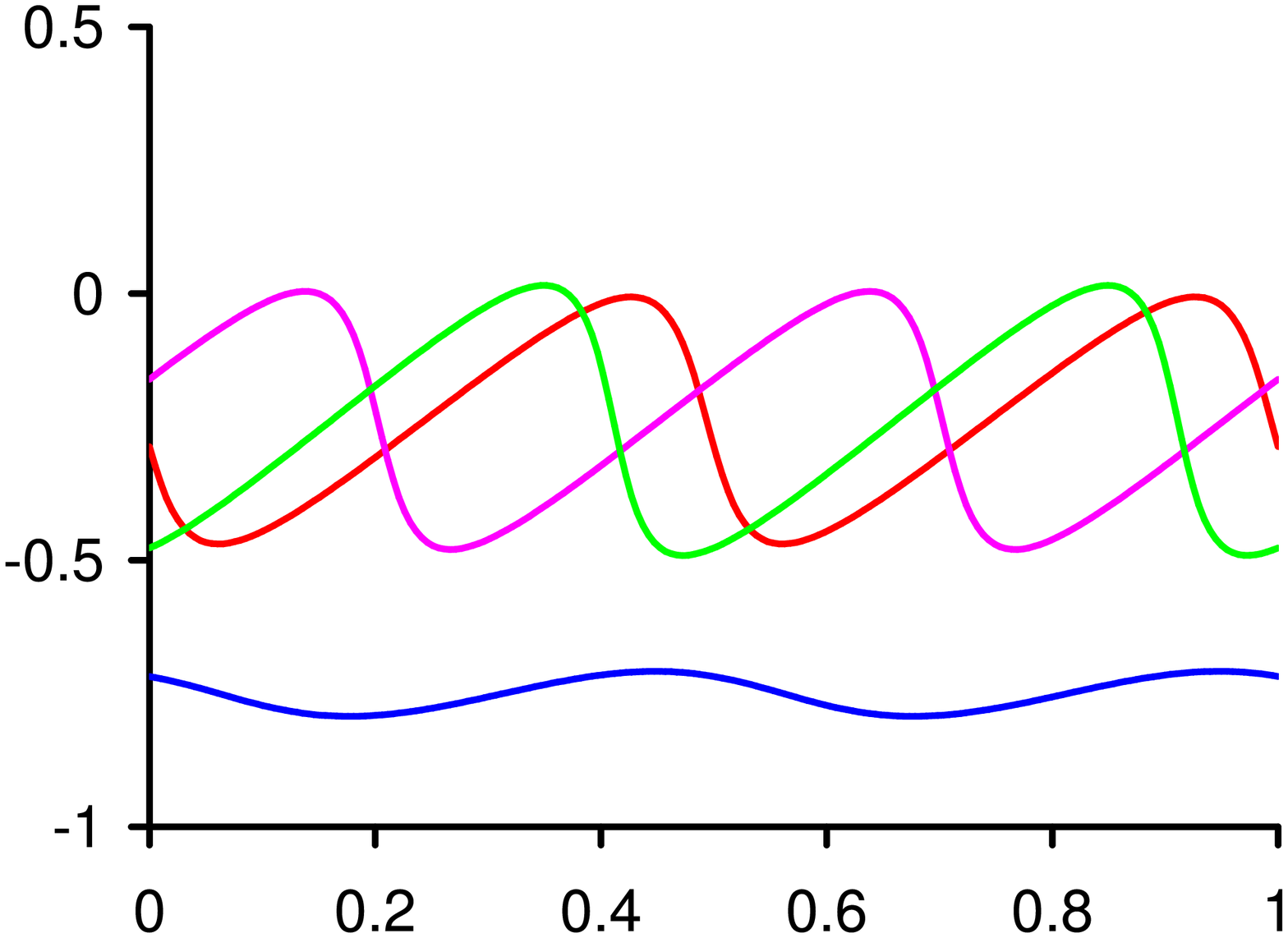}}}
\put(80,0){Relative cavity length}
\put(0,170){Phase/$\pi$}}
\end{picture}
\caption{\label{figFBTransAndPhase} Transmission (left) and phase (right) of light passing through a Fabry-P\'{e}rot cavity A to D as shown in figure \ref{figFBCavity} as function of the relative cavity length (this is the fractional about of the cavity length expressed in wavelength). The blue line (No) gives the same quantities when the absorption is ignored.}
\end{figure}

With two semi-transparent mirrors a Fabry-P\'{e}rot cavity can be constructed. Here, the semi-transparent mirrors are made out of a thin single layer of metal (25 nm of Silver) on a piece of glass (BK7) as schematically shown in figure~\ref{figFBCavity}. The mirrors can be positioned in four different ways to form the cavity: (A) glass substrates facing each other, (B) and (C) glass substrate facing metal layer and (D) metal layers facing each other. The length of the cavity can be varied. Normally the transmission of the cavity is maximized and the finesse should be very high. Here, the finesse should not be too high, as some light must be transmitted for the measurement. Hence, the metal layers are sufficiently thick to absorb part of the light and sufficiently thin not to be opal. 
Here, a thickness of approximately 25 nm was chosen and a wavelength of 0.633~$\mu$m, corresponding to the wavelength of a Helium-Neon laser. The results for the transmission and phase of a light beam passing through a cavity is shown in figure~\ref{figFBTransAndPhase} for the situations A to D as described above. For each cavity the transmission has a maximum value when the standing wave matches the cavity. In that case the transmission is much larger than in the case for a single layer (values shown in figure~\ref{figAgSamples} for 0.633~$\mu$m). The minimum value is of the order of 1.5~\% corresponding to the transmission through two single layers. The phase dependence shows that the phase of the light exiting the cavity behaves a-typical. When the transmission is low, the phase decreases when the cavity length is increased. When the transmission is high, the phase increases when the cavity length is increased. Note that the phase is relative to that of a light beam traveling the same distance through air. When the cavity would be made of non-absorbing materials only, the phase would be almost constant as shown by the blue lines. Almost, as also the multiple reflections inside the cavity influence the phase behavior, although to a much lesser extend. If also the reflection would be ignored the transmission becomes 1 and the phase constant.
\begin{figure}
\begin{picture}(400,180)
\put(0,0){
\put(0,15){\scalebox{0.35}{\includegraphics{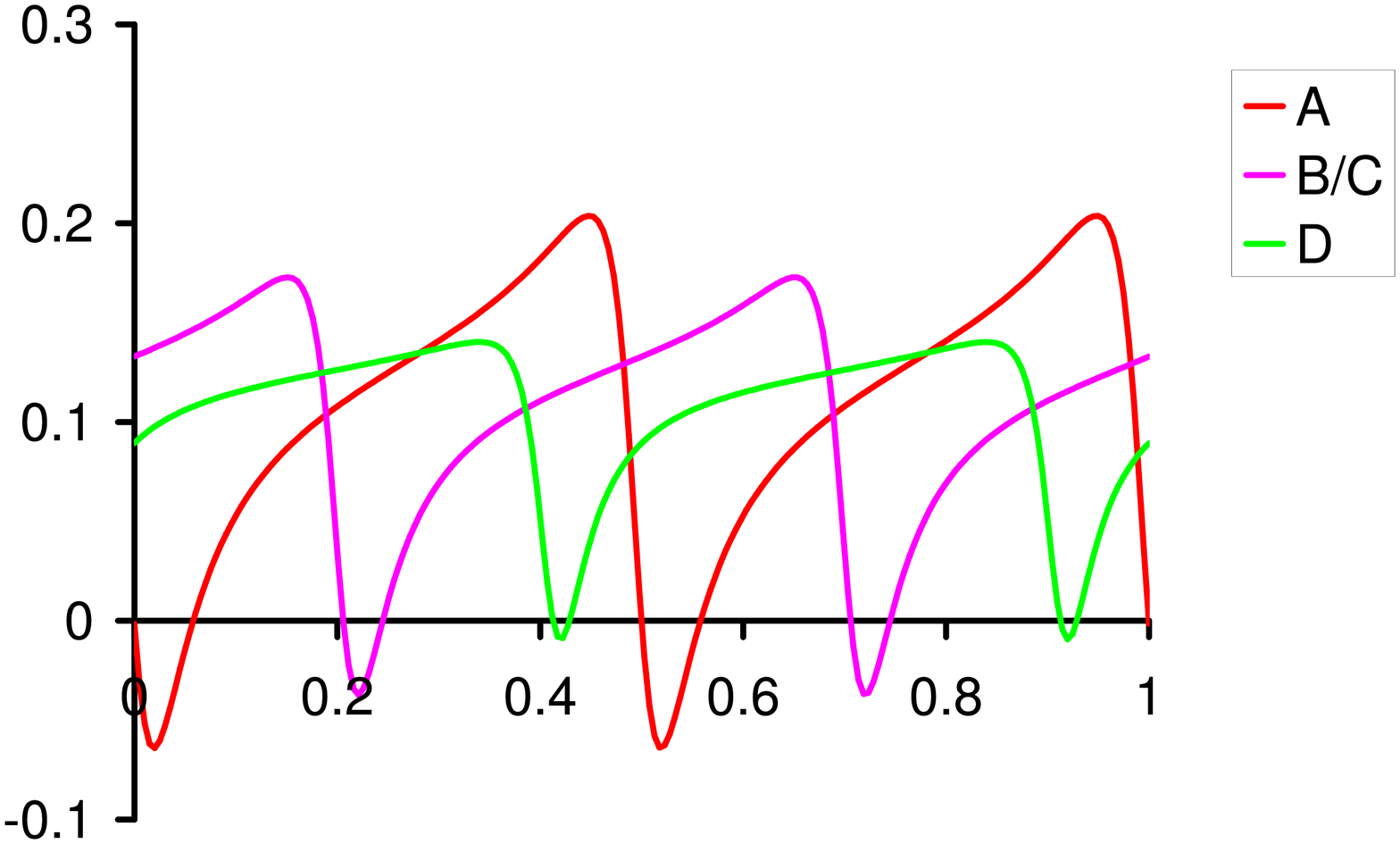}}}
\put(80,0){Relative cavity length}
\put(0,170){Sensitivity in rad}}
\put(240,0){
\put(0,15){\scalebox{0.35}{\includegraphics{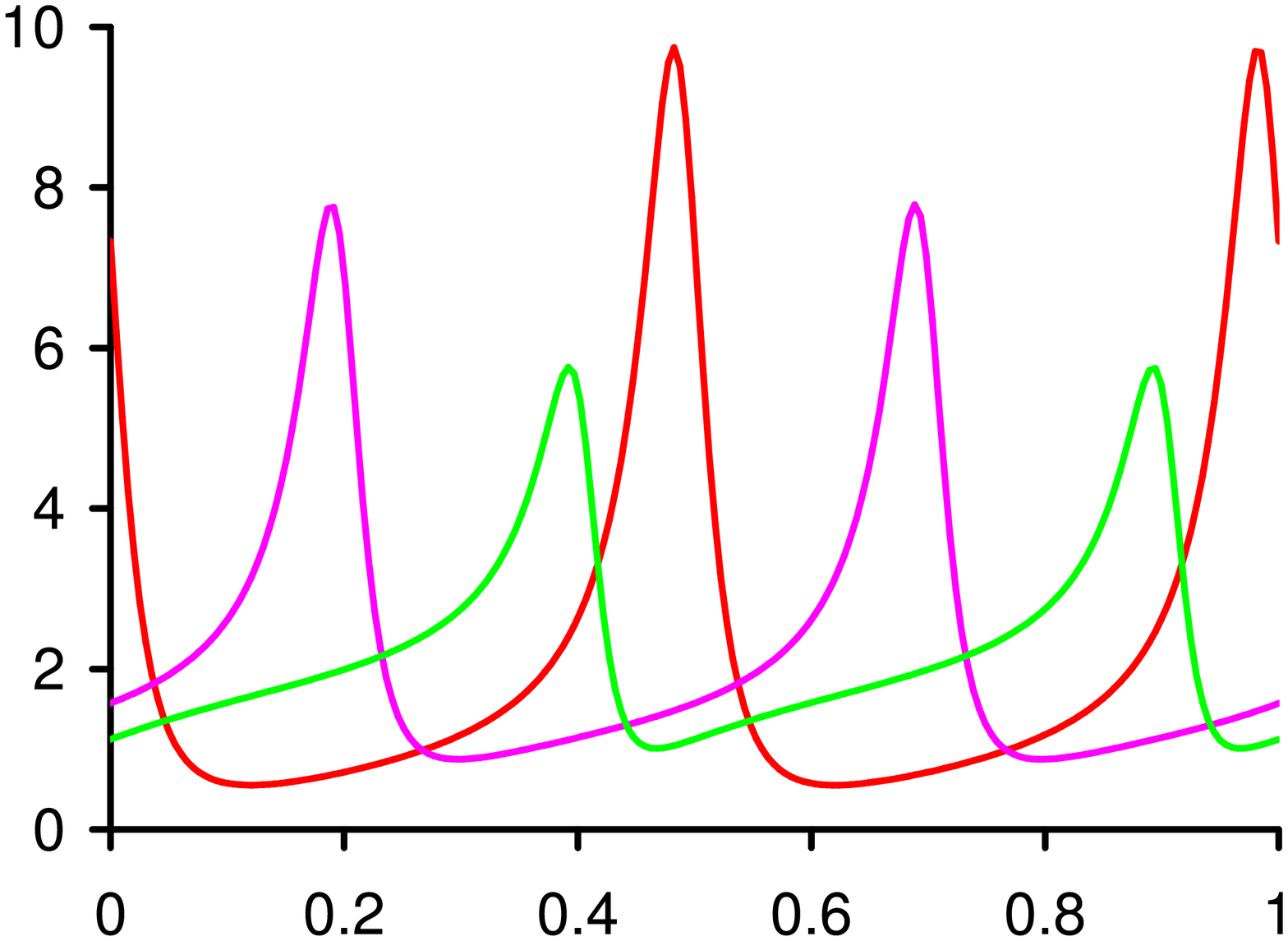}}}
\put(80,0){Relative cavity length}
\put(0,170){Sensitivity in rad}}
\end{picture}
\caption{\label{figFBSensitivity} Sensitivity, $S_\sigma$ as function of the cavity length of Fabry-P\'{e}rot cavity A to D as shown in figure \ref{figFBCavity} as function of the relative cavity length. Left: sensitivity due to conductivity changes; Right: sensitivity due to plasma frequency changes.}
\end{figure}

When for a fixed cavity length the conductivity of the Silver layers changes, also the transmission and the phase change. The sensitivity, $S_\sigma$ as function of the cavity length is shown in the left graph of figure~\ref{figFBSensitivity} for cavity types A to D. The maximum sensitivity occurs when the transmission is lowest. When somehow the conduction is constant, but the plasma frequency changes, the sensitivity is shown in the right graph of figure~\ref{figFBSensitivity}.

\section{Mach-Zehnder interferometer} \label{secMZ}

\begin{figure}[t]
\begin{picture}(180,180)
\put(20,0){\scalebox{0.8}{\includegraphics{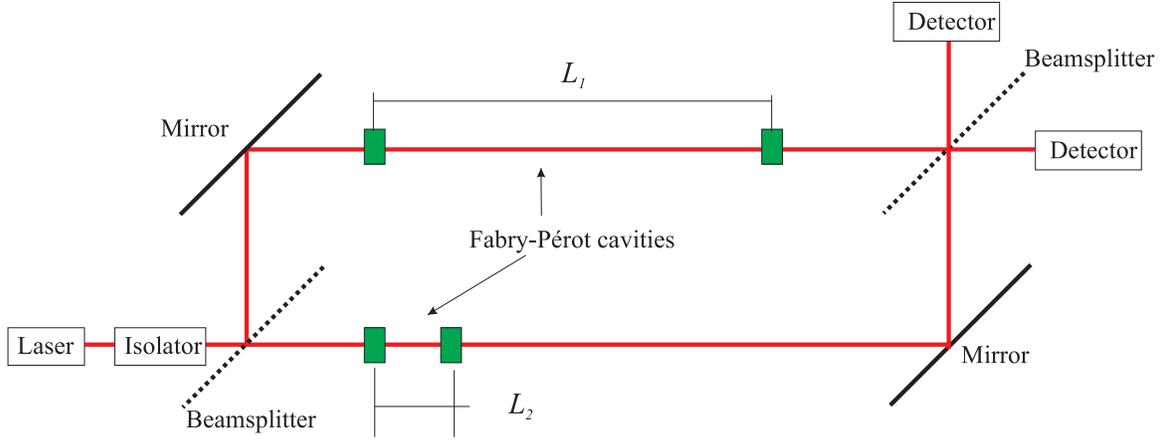}}}
\end{picture}
\caption{\label{figMZ} Mach-Zehnder geometry for double Fabry-P\'{e}rot cavity.}
\end{figure}

If such Fabry-P\'{e}rot cavities are put in each arm of an optical (wavelength, $\lambda$) Mach-Zehnder interferometer (see figure~\ref{figMZ}), then the output signal is determined by the intensity and phase difference between the two optical beams. 
\begin{figure}[b]
\begin{picture}(180,100)
\put(140,0){\scalebox{0.6}{\includegraphics{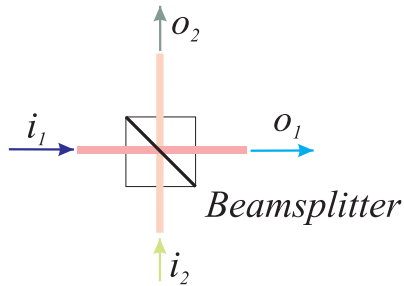}}}
\end{picture}
\caption{\label{figBeamSplitter} Sketch of beam splitter geometry.}
\end{figure}

The function of the beam splitters can be described by using the transfer matrix approach~\cite{Chen2007},\cite{deHaan2011}. Let $i_1$ and $i_2$ be the amplitudes of the light waves at the two inputs of the beam splitter and $o_1$ and $o_2$ the same for the outputs (see figure~\ref{figBeamSplitter}), then for an ideal beam splitter
\[
\left( \begin{array}{c} o_1  \\ o_2 \end{array} \right)=\frac{1}{\sqrt{2}}
\left( \begin{array}{cc} 1 & -i  \\ -i & 1 \end{array} \right) 
\left( \begin{array}{c} i_1  \\ i_2 \end{array} \right) 
\]
The 2$\times 2$~matrix is referred to as {\it transfer matrix} and describes the action of the beam splitter. The incident light wave present at the first beam splitter, $i_1$ is provided by a He-Ne laser with a wavelength of 632.8~nm. When only one light beam is applied, the output of both exit waveguides can be inferred by taking $i_1=1$ and $i_2=0$. Then according to the above equation
\[
\left( \begin{array}{c} o_1  \\ o_2 \end{array} \right)=\frac{1}{\sqrt{2}}\left( \begin{array}{cc} 1  \\ -i  \end{array} \right) 
\]
The wave energy is split over both exits. The phase difference between the waves of both exits is $-\pi/2$. The propagation of the wave through the arms of the interferometer passing the Fabry-P\'{e}rot cavity until the second beam splitter can be described by a simple multiplication of the amplitude at the beginning of the arm by the transmission and a phase factor describing the optical path length experienced by the light. Here, it is assumed that the transmission of the first arm is $\tau_1$ (complex to take into account the phase difference) and $\tau_2$ for the second arm. At the second beam splitter the waves interfere and result in the waves at the exits denoted by $o_3$ and $o_4$ according to equation the above equation where $i_1$ and $i_2$ are replaced by $i_3$ and $i_4$ and similar replacement for $o_1$ and $o_2$
\[
\left( \begin{array}{c} o_3  \\ o_4 \end{array} \right)=\frac{1}{\sqrt{2}}\left( \begin{array}{cc} 1 & -i  \\ -i & 1 \end{array} \right)\left( \begin{array}{c} i_3  \\ i_4 \end{array} \right) 
\]
If the above equations are combined one gets
\[
\left( \begin{array}{c} o_3  \\ o_4 \end{array} \right)=\frac{1}{2}\left( \begin{array}{cc} 1 & -i  \\ -i & 1 \end{array} \right) \left( \begin{array}{cc} \tau_1 & 0  \\ 0 & \tau_2 \end{array} \right) \left( \begin{array}{cc} 1  \\ -i  \end{array} \right)  
\]
The sum of the intensities at the exit of the last beam splitter can be written as
\[
\left|o_{4}\right|^2+\left|o_{3}\right|^2 = \frac{T_1+T_2}{2}
\]
where $T_1=|\tau_1|^2$ is the transmission of the Fabry-P\'{e}rot cavity in arm 1 and a similar relation for arm 2. The transmission of both cavities can vary with the length of the cavity according to figure~\ref{figFBTransAndPhase} depending of the type A to D. When the glass thickness varies one has to take that into account the by an appropriate averaging.
The visibility of this interferometer can be defined as the relative difference between the two outputs of the last beam splitter, hence
\[
V =  \frac{\left|o_{4}\right|^2-\left|o_{3}\right|^2}{\left|o_{4}\right|^2+\left|o_{4}\right|^2} =\frac{2\sqrt{T_1T_2}}{T_1+T_2}\cos(\phi_2-\phi_1) 
\]
where $\phi_1=\arg \tau_1$ is the phase acquired by the light beam while passing through arm 1 and a similar relation for arm 2. Here also, the phases $\phi_1$ and $\phi_2$ vary with the same dependence as the transmissions.

\begin{figure}[h]
\begin{picture}(450,140)
\put(15,0){
\put(390,5){\put(22,8){0}\put(22,35){0.5}\put(22,64){1}}
\put(390,5){\scalebox{0.40}{\includegraphics{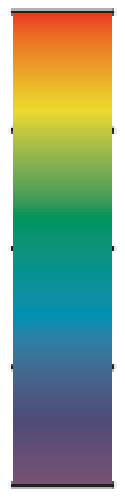}}}
\put(10,10){\scalebox{0.60}{\includegraphics{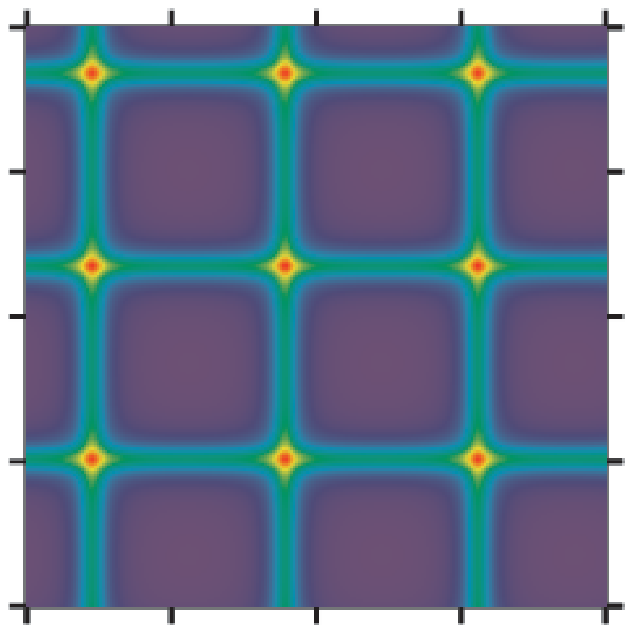}}}
\put(0,5){
\put(0,6){$20$}
\put(0,56){$35$}
\put(0,107){$50$}
\put(10,-5){$20$}
\put(64,-5){$35$}
\put(110,-5){$50$}
\put(0,122){$V_1$}
\put(85,-10){$V_2$}
}
\put(140,10){\scalebox{0.60}{\includegraphics{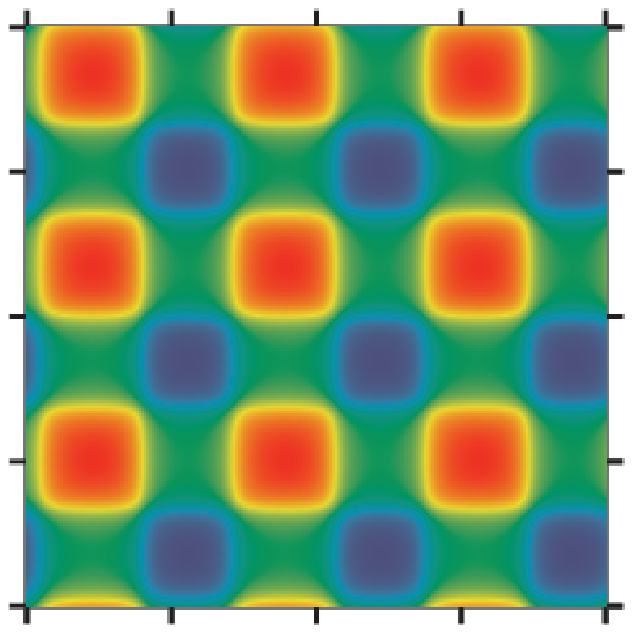}}}
\put(130,5){
\put(0,6){$20$}
\put(0,56){$35$}
\put(0,107){$50$}
\put(10,-5){$20$}
\put(64,-5){$35$}
\put(110,-5){$50$}
\put(0,122){$V_1$}
\put(85,-10){$V_2$}
}
\put(270,10){\scalebox{0.60}{\includegraphics{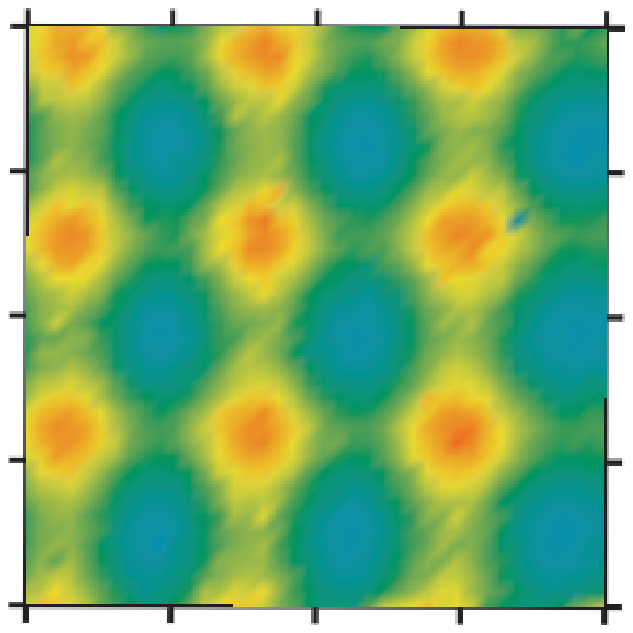}}}
\put(260,5){
\put(0,6){$20$}
\put(0,56){$35$}
\put(0,107){$50$}
\put(10,-5){$20$}
\put(64,-5){$35$}
\put(110,-5){$50$}
\put(0,122){$V_1$}
\put(85,-10){$V_2$}
}}
\end{picture}
\caption{\label{figSumMZ} Theoretical ideal (Left), theoretical non-ideal (Middle) and measured (Right) sum of intensities of outputs of a Mach-Zehnder interferometer as function of the voltages applied to piezo stacks.}
\end{figure}

\begin{figure}
\begin{picture}(450,140)
\put(15,0){\put(390,5){\put(22,8){0}\put(22,35){1}\put(22,64){2}}
\put(390,5){\scalebox{0.40}{\includegraphics{FigScale2Dplot.eps}}}
\put(10,10){\scalebox{0.60}{\includegraphics{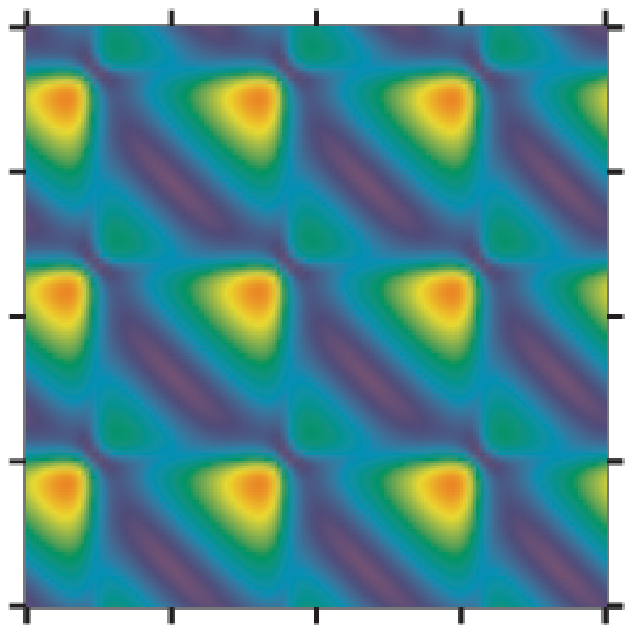}}}
\put(0,5){
\put(0,6){$20$}
\put(0,56){$35$}
\put(0,107){$50$}
\put(10,-5){$20$}
\put(64,-5){$35$}
\put(110,-5){$50$}
\put(0,122){$V_1$}
\put(85,-10){$V_2$}
}
\put(140,10){\scalebox{0.60}{\includegraphics{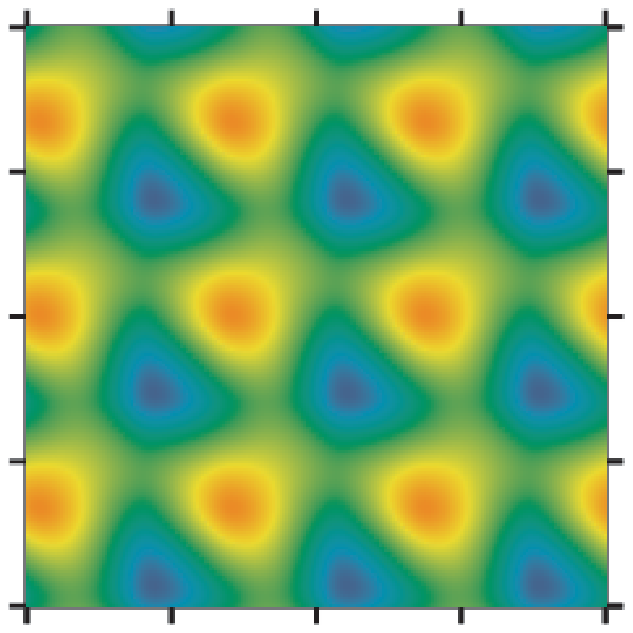}}}
\put(130,5){
\put(0,6){$20$}
\put(0,56){$35$}
\put(0,107){$50$}
\put(10,-5){$20$}
\put(64,-5){$35$}
\put(110,-5){$50$}
\put(0,122){$V_1$}
\put(85,-10){$V_2$}
}
\put(270,10){\scalebox{0.60}{\includegraphics{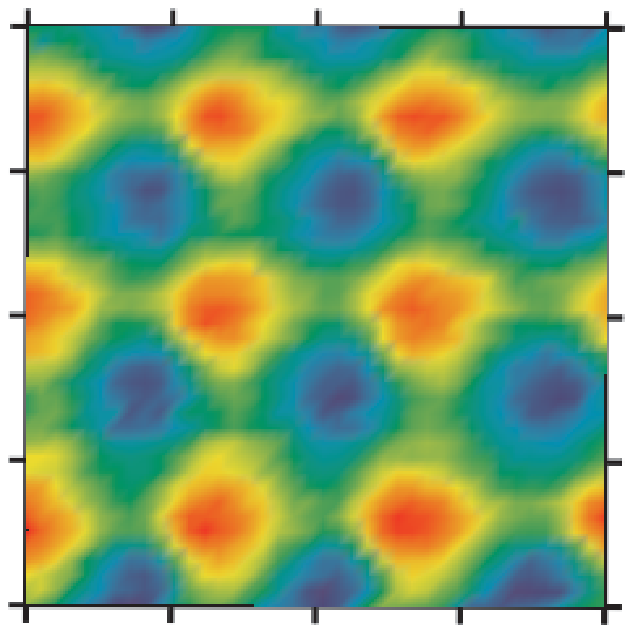}}}
\put(260,5){
\put(0,6){$20$}
\put(0,56){$35$}
\put(0,107){$50$}
\put(10,-5){$20$}
\put(64,-5){$35$}
\put(110,-5){$50$}
\put(0,122){$V_1$}
\put(85,-10){$V_2$}
}}
\end{picture}
\caption{\label{figPhaseMZ} 
Theoretical ideal (Left), theoretical non-ideal (Middle) and measured (Right) phase difference in radian of a Mach-Zehnder interferometer as function of the voltages applied to piezo stacks.}
\end{figure}

To check if the above described effects are observable in an optic interferometer, measurements were performed with the Mach-Zehnder interferometer corresponding to figure~\ref{figMZ} with Fabry-P\'{e}rot cavities as described in the previous section. The light of a stabilized He-Ne laser (type Coherent 200, linear polarized, 0.5 mW, maximum mode sweep 10 MHz) was coupled into the $i_1$ arm of the interferometer via a polarization dependent optical isolator (isolation at least 35 dB). The interferometers were put in a temperature controlled environment where the temperature control was within 3~mK. 
One side of each Fabry-P\'{e}rot cavity was mounted by means of a piezo stack enabling the independent change of the length of the cavities. The light intensity at outputs $o_{3}$ and $o_{4}$ of the Mach-Zehnder interferometer were measured by two amplified silicon detectors. The sum of these intensities relative to the output intensity of the laser as function of the voltage applied to the piezo stacks is given in the right graph of figure~\ref{figSumMZ}. For comparison also the ideal theoretical values are shown in the left graph. The occurrence of minima and maxima is the same, but the sharpness of the features in the measurements is much less than for the ideal theoretical calculations. This is due to the uncertainties on the exact sample parameters. Especially the glass thickness variations play an important role as is shown in the middle graph of figure~\ref{figSumMZ}. This graph was calculated by using a variation in the glass thickness of $\pm  20$~nm which is comparable to the surface flatness of the substrate used (65~nm).
The visibilities can be calculated from the intensities at outputs $o_{3}$ and $o_{4}$ according to the above equation. They are shown in the right graph of figure~\ref{figPhaseMZ} as function of the voltage applied to the piezo stacks. For comparison also the ideal theoretical values are shown in the left graph. Here also the occurrence of minima and maxima is the same, but the sharpness of the features in the measurements is much less than for the ideal theoretical calculations. The influence of the same glass thickness variations as before result in the middle graph of figure~\ref{figPhaseMZ}. 
For increasing $V_1$ voltage it seems that the minima and maxima shift a bit to larger values of voltage $V_2$. During the measurements voltage $V_1$ was set to a fixed value and $V_2$ was changed from its minimum to maximum value. A small drift in the interferometer path length of one arm with respect to the other will result in this small drift (a complete measurement took about 80 min).
The measurements show that the transmission of the Fabry-P\'{e}rot cavities behave as expected and that it is possible to measure the phase changes due to variation of the properties of the cavities. This behavior is exploited in the experiment that is described in the next section.

\section{Sun and Earth velocity components}
Let us define the Sun and Earth velocity components as measured by an observer on Earth. The rotation speed of the Earth around its own axis will be ignored.

\begin{figure}
\begin{picture}(450,220)
\put(0,0){\scalebox{0.50}{\includegraphics{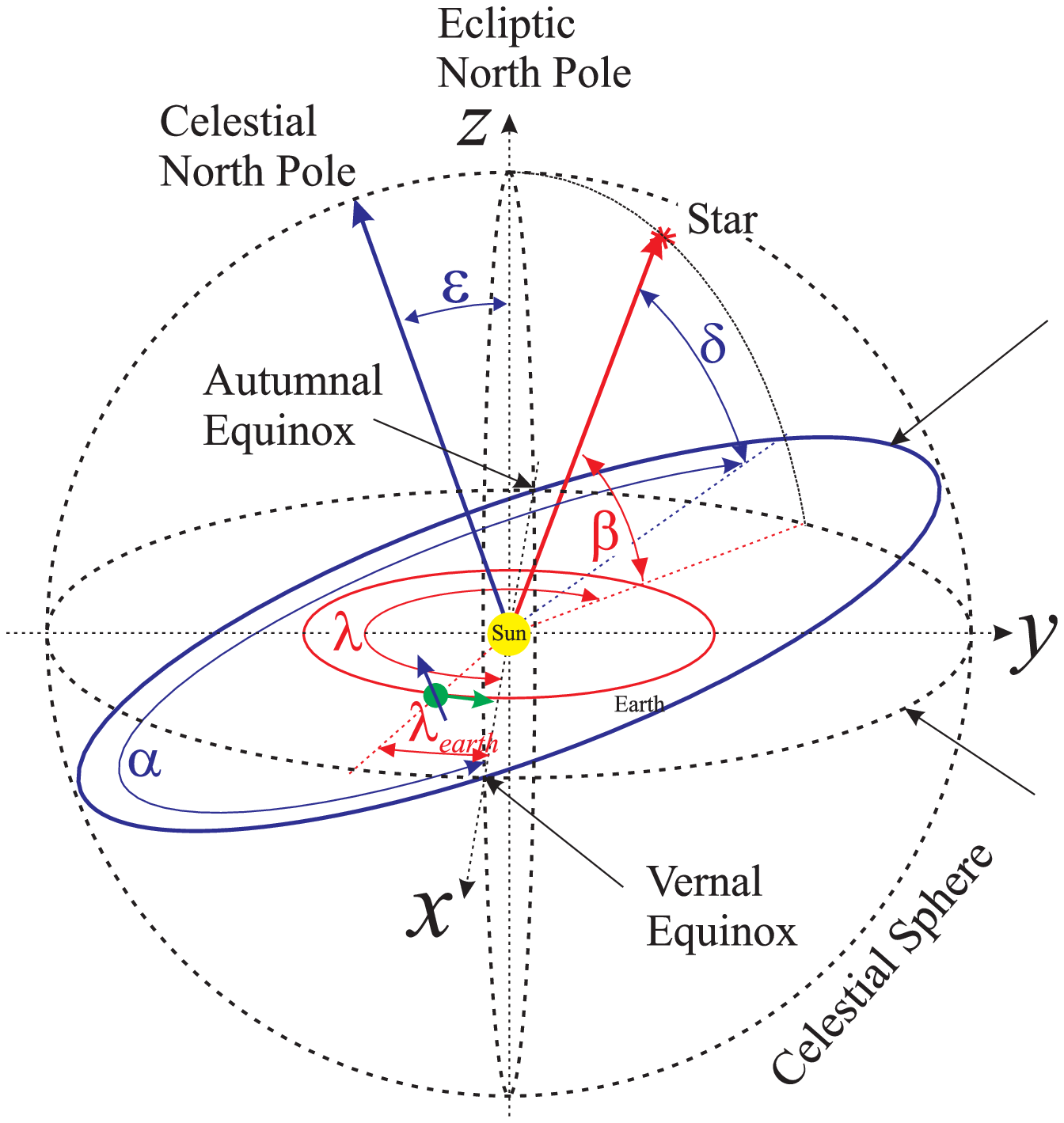}}}
\put(260,0){\scalebox{0.50}{\includegraphics{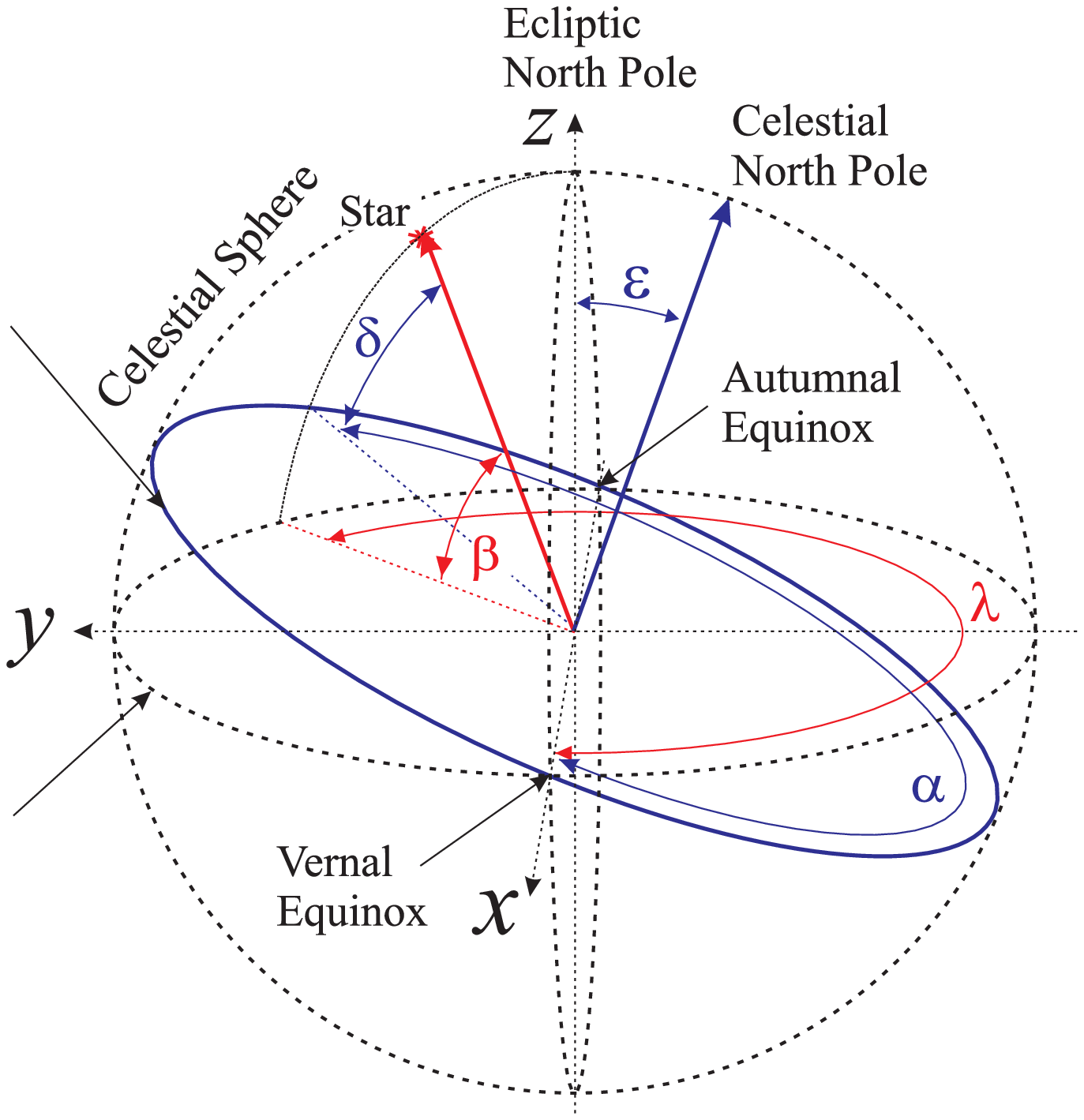}}}
\put(170,130){
\put(20,22){\small Celestial Equator}
\put(20,10){\small (projection of}
\put(20,-2){\small Earth's equator)}
}
\put(180,35){
\put(20,24){\small Ecliptic}
\put(20,12){\small (projection of}
\put(20,0){\small Earth's orbit)}
}
\end{picture}
\caption{\label{figCelestial} 
Definition of celestial coordinate systems. Left: Celestial view; Right: Earth view. See the text for the explanation of the symbols. 
}
\end{figure} 

Figure~\ref{figCelestial} shows the definition of the different coordinate systems and their connections. Two views are presented. The left view represents the celestial view with the Sun in the origin. The right view represents the Earth view with the Earth in the origin. The ecliptic or celestial coordinates are defined with respect to the Earth orbit around the Sun. $\lambda$ is the ecliptic or celestial longitude (red angle from vernal equinox). $\beta$ is the ecliptic of celestial latitude (red angle from plane through ecliptic). The so-called {\it fixed stars} can be found in a direction for which $\beta$ and $\lambda$ are constant. In vector notation the direction of the fixed stars is given in ecliptic coordinates by
\[
\vec{R}_C=\left(\begin{array}{c} \cos\beta \cos\lambda \\ -\cos\beta \sin\lambda \\ \sin \beta \end{array}\right)
\]
The equatorial coordinates are defined with respect to the projection of the Earth equator. This projection makes an angle $\phi_{tilt}$ with the ecliptic. $\phi_{tilt}$ is the Earth's axial tilt with respect to the ecliptic. The rotation axis is the line between the vernal and autumnal equinox. In the coordinate system used, this corresponds to the $x$-axis. Currently $\phi_{tilt}=23.44^o$, but the Earth axis processes slowly so that the equatorial coordinates only have meaning when the epoch of observation is given. 
$\alpha$ is the right ascension (blue angle along celestial equator) and $\delta$ is the declination (blue angle going up from plane through celestial equator). In vector notation the direction of the fixed stars is given in ecliptic coordinates by
\[
\vec{R}_E = \left(\begin{array}{c} \cos\delta \cos\alpha \\ -\cos\delta \sin\alpha \\ \sin \delta \end{array}\right)
\]
It is possible to transform the Equatorial coordinates into Celestial coordinates and vice versa by means of
\[
\vec{R}_C=  \left(\begin{array}{ccc} 1&0&0\\ 0&\cos\phi_{tilt}& -\sin\phi_{tilt} \\ 0 & \sin\phi_{tilt} & \cos\phi_{tilt} \end{array} \right)\vec{R}_E  \ \ \ \ \ {\rm and }\ \ \ \ \ \vec{R}_E =  \left(\begin{array}{ccc} 1&0&0\\ 0&\cos\phi_{tilt}& \sin\phi_{tilt} \\ 0 & -\sin\phi_{tilt} & \cos\phi_{tilt} \end{array} \right)\vec{R}_C
\]
\begin{figure}
\begin{picture}(450,160)
\put(0,0){\scalebox{0.50}{\includegraphics{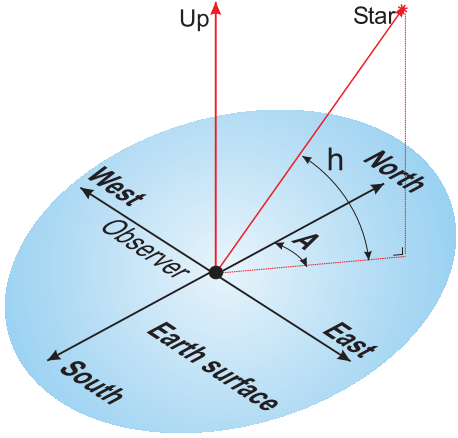}}}
\end{picture}
\caption{\label{figEarth} 
Definition of the coordinate system of the observer at some position on Earth (Horizontal coordinate system). See the text for the explanation of the symbols. 
}\end{figure} 
With respect to an observer on Earth, the coordinate system used is shown in figure~\ref{figEarth}. The Earth coordinate system is defined with respect to the local North. $A$ is the Azimuth (black angle on Earth surface from North direction toward East. $h$ is the elevation (black angle going up from Earth's surface). 
In vector notation a direction in Earth coordinates is given by
\[
\vec{R}_A = \left(\begin{array}{c} \cos h \cos A \\ -\cos h \sin A \\ \sin h \end{array}\right)
\]
When the observer's location on Earth is defined by its longitude ($\lambda_O$) and latitude ($\phi_O$), it is possible to transform the Earth coordinates into Equatorial coordinates and vice versa depending on the time (universal time $t_{UT}$) and date of observation by means of
\[
\vec{R}_E =  \left(\begin{array}{ccc} -\sin \phi_O \cos \lambda & \sin \lambda & \cos\phi_O\cos \lambda\\ \sin\phi_O\sin\lambda &\cos\lambda & -\cos\phi_O\sin\lambda \\ \cos\phi_O & 0 & \sin \phi_O \end{array} \right)\vec{R}_A\]
and
\[ \vec{R}_A =  \left(\begin{array}{ccc} -\sin \phi_O \cos \lambda & \sin \lambda \sin\phi_O & \cos\phi_O \\ \sin\lambda &\cos\lambda & 0 \\ \cos\lambda \cos\phi_O & -\cos\phi_O\sin\lambda & \sin \phi_O \end{array} \right)\vec{R}_E
\]
where $\lambda = \lambda_O + \lambda_{Earth} + 2\pi t_{UT}/T_{Day} $ is related to the sidereal time by
\[
t_{sid}=T_{Day} \frac{|\lambda-\pi,2\pi|}{2\pi}
\]
where the function $|a,b|$ denotes the value of $a$ modulo $b$. 

The ecliptic longitude of the Earth at observation time is given by 
\[
\lambda_{Earth} = 2\pi \frac{\Delta T_{SE}}{T_{Orbit}}
\]
where $T_{Orbit}=365.26$ days is the orbit time of the Earth around the Sun. $\Delta T_{SE}$ is the time passed since the moment the Earth was at the latest vernal (or spring) equinox. It can be calculated by means of the Julian Date, $JD$ by 
\[
\Delta T_{SE} = \left|JD+343.85743,T_{Orbit}\right|
\]
where $JD$ is the count of days since noon Universal Time on January 1, 4713 BCE (on the Julian calendar). Hence, the Julian Date of January 1, 2000 12h UT equals 2451545. The Julian date can be inferred from the Julian calender ($DD/MM/YYYY$) by means of
\[
JD = DD +30MM+367YYYY+1721029 - 2Q + F\left(\frac{7(MM-2)}{12}\right) + F\left(\frac{Q}{4}\right)-F\left(\frac{3P}{4}\right)
\]
where
\[
Q=YYYY+F\left(\frac{MM-3}{12}\right)  \ \ \ \ {\rm and } \ \ \ \ P = 1 + F\left(\frac{Q}{100}\right)
\]
and $F(x)$ denotes the largest integer below $x$.

Let us define the velocity of the Sun through the Ether in ecliptical coordinates as 
\[
\vec{v}_{Sun,C}=v_{Sun}\left(\begin{array}{c} \cos\beta_{Sun} \cos\lambda_{Sun}  \\ -\cos\beta_{Sun} \sin\lambda_{Sun}  \\ \sin \beta_{Sun} \end{array}\right)
\]
where $\beta_S$ is the ecliptical latitude and $\lambda_{Sun} $ is the ecliptical longitude and $v_{Sun}$ is the magnitude of Sun's velocity through the ether. In the equatorial coordinate system this becomes
\[
\vec{v}_{Sun,E}= v_{Sun}\left(\begin{array}{c} \cos\beta_{Sun} \cos\lambda_{Sun}  \\ \sin\phi_{tilt}\sin\beta_{Sun}  -\cos\phi_{tilt} \cos\beta_{Sun} \sin\lambda_{Sun}  \\ \sin\phi_{tilt}\cos\beta_{Sun} \sin\lambda_{Sun}  + \cos\phi_{tilt} \sin \beta_{Sun} \end{array}\right) 
\]
The velocity of the Earth with respect to the Sun in Ecliptical coordinated is given by
\[
\vec{v}_{Earth,C}=v_{Earth}\left(\begin{array}{c}  \sin \lambda_{Earth} \\ \cos\lambda_{Earth} \\ 0\end{array}\right)
\]
or in equatorial coordinates
\[
\vec{v}_{Earth,E}= v_{Earth} \left(\begin{array}{c}  \sin \lambda_{Earth} \\ \cos\phi_{tilt} \cos\lambda_{Earth} \\ -\sin\phi_{tilt}\cos\lambda_{Earth} \end{array}\right)
\]
where $v_{Earth}=29.78$ km/s. Hence, the velocity of the observer with respect to the ether is $\vec{v}=\vec{v}_{Sun,C}+\vec{v}_{Earth,C}$, where the rotation of the Earth is ignored. This velocity can be transformed in the Earth coordinate system by applying the above transformation. 

The velocity of the Sun with respect to the ether is to be determined from the measurements. Miller~\cite{Miller1933} in his repetitions of the Michelson-Morley experiment found a preferred direction from his measurements, the values of which are shown in table~\ref{tabSun}.  Miller 1 refers to his first derivation and Miller 2 to his more accurate determination in 1933. Both directions are nearly opposites. Another possible candidate is the reference frame in which the microwave background radiation is uniform (CMB)~\cite{Smoot1977}. This direction is almost perpendicular to the first two mentioned.

\begin{table}
\begin{tabular}{|l|l|l|l|l|l|l|} \hline
Parameter            & Symbol          & Unit    & CMB   &  Miller 1 & Miller 2  \\ \hline
Ecliptical latitude  & $\beta_{Sun}$   & Degree  & 171.9 &     84    & -83       \\ \hline 
Ecliptical longitude & $\lambda_{Sun}$ & Degree  & -11.1 &    159    & -41       \\ \hline
Right ascension      & $\alpha_{Sun}$  & Degree  & 168.2 &    255    &  73.5     \\ \hline
Declination          & $\delta_{Sun}$  & Degree  &  -7.0 &     68    & -70.6     \\ \hline
Magnitude velocity   & $v_{Sun}$       & km/s    & 369   &    205    &  205      \\ \hline
\end{tabular}
\caption{\label{tabSun} 
Sun velocity parameters for several models.}
\end{table}

\section{Effect}
Let us assume that the influence of the absolute velocity of the material with respect to the ether, $\vec{v}$ on the effective conductivity and plasma frequency is given by
\[
\sigma_v = \sigma \left(1+\eta_\sigma \frac{\vec{k}\cdot\vec{v}}{kc} \right)
\]
where $|\eta_\sigma|$ is of the order of 1. The value of $\eta_\sigma$ depends on the way that the conductivity depends on the velocity with respect to the ether. For now this is unknown, but it should not differ too much from 1. 
Let $\phi$ be the phase a light beam acquires when traversing through the Fabry-P\'{e}rot cavities as defined in the previous sections. Then, the phase difference between phases acquired by a wave traversing when the direction of the light beam through the cavities changes from $\vec{\zeta}_o$ to $\vec{\zeta}$ (both unit vectors) with respect to $\vec{v}$ is
\[
\Delta \phi = S_{\sigma} \eta_\sigma \frac{(\vec{\zeta}-\vec{\zeta_o})\cdot\vec{v}}{c} 
\]
When the interferometer is rotated along an axis represented by unit vector $\vec{m}$, with an angle $\theta$ the vector $\vec{\zeta}$ is given by the Rodriguez rotation formula
\[
\vec{\zeta} = \vec{\zeta}_o + \vec{m}\times\vec{\zeta}_o \sin \theta - \left(\vec{\zeta}_o-\vec{m}(\vec{m}\cdot\vec{\zeta}_o)\right)(1-\cos\theta)
\]
rotating the vector $\vec{\zeta}_o$ around a unit vector $\vec{m}$ over an angle $\theta$ according to the right hand rule. When $\vec{\zeta}_o \perp \vec{m}$ then this reduces to
\[
\vec{\zeta} = \vec{\zeta}_o \cos \theta + \vec{\zeta}_{\perp} \sin \theta 
\]
where $\vec{\zeta}_{\perp}= \vec{m}\times\vec{\zeta}_o $ is in the plane of rotation perpendicular $\vec{\zeta}_o$. Hence,
\[
\Delta \phi = S_{\sigma} \eta_\sigma \hat{\beta} \left( \cos (\theta - \theta_o)  - \cos \theta_o \right) 
\]
where $\hat{\beta}=|\vec{v}\times\vec{m}|/c$ the magnitude of the component of the ether velocity vector in the plane of rotation relative to the speed of light and $\cos \theta_o = \vec{\zeta}_o\cdot\vec{v}/|\vec{v}\times\vec{m}|$, the length of that component parallel to $\vec{\zeta}_o$. Hence, for a complete rotation $\Delta \phi$ changes according to a cosine with an amplitude of $\hat{\beta}$. The maximum occurs when $\theta=\theta_o$ and the minimum in the opposite direction. Both $\hat{\beta}$ and $\theta_o$ depend on the magnitude and direction of the ether velocity with respect to the rotation plane and the initial direction of the setup.    

The normal to the rotation plane can be expressed in the Earth's coordinate system according to
\[
\vec{m}_A=  \left(\begin{array}{c} 0 \\ 0 \\ 1 \end{array}\right)
\]
so that in the equatorial coordinate system
\[
\vec{m}_E =  \left(\begin{array}{ccc} -\sin \phi_O \cos \lambda & \sin \lambda & \cos\phi_O\cos \lambda\\ \sin\phi_O\sin\lambda &\cos\lambda & -\cos\phi_O\sin\lambda \\ \cos\phi_O & 0 & \sin \phi_O \end{array} \right)\left(\begin{array}{c} 0 \\ 0 \\ 1 \end{array}\right)=\left(\begin{array}{c} \cos\phi_O\cos \lambda \\ -\cos\phi_O\sin \lambda \\  \sin \phi_O\end{array}\right)
\]
Let us assume that $\zeta_o$ is in the South-North direction so that its Azimuth is 0, hence in Earth coordinate system
\[
\vec{\zeta}_{o,A} =  \left(\begin{array}{c} 1 \\ 0 \\ 0 \end{array}\right)
\]
so that in the equatorial coordinate system
\[
\vec{\zeta}_{o,E} = \left(\begin{array}{c}  -\sin \phi_O \cos \lambda \\ \sin\phi_O\sin\lambda \\ \cos\phi_O \end{array}\right)
\]
From these equations the maximum magnitude and the direction of the maximum can be determined.
\begin{figure}
\begin{picture}(450,200)
\put(0,0){\scalebox{1}{\includegraphics{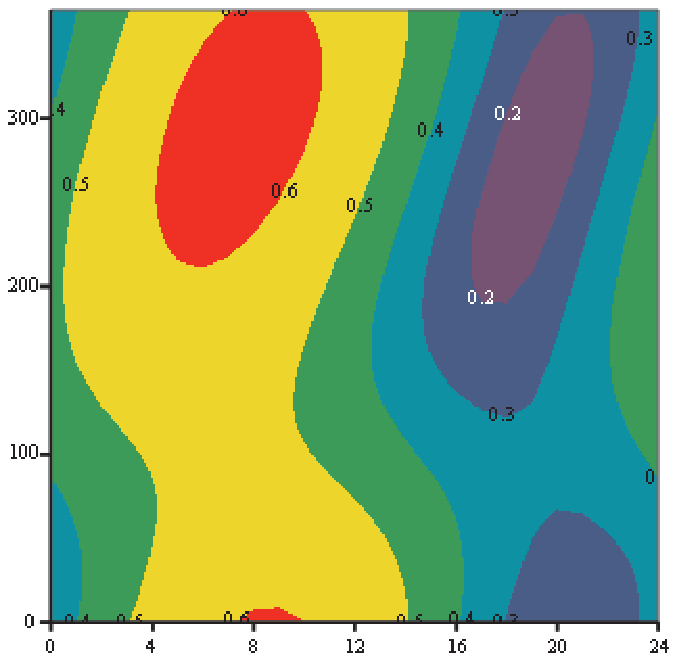}}}
\put(250,0){\scalebox{1}{\includegraphics{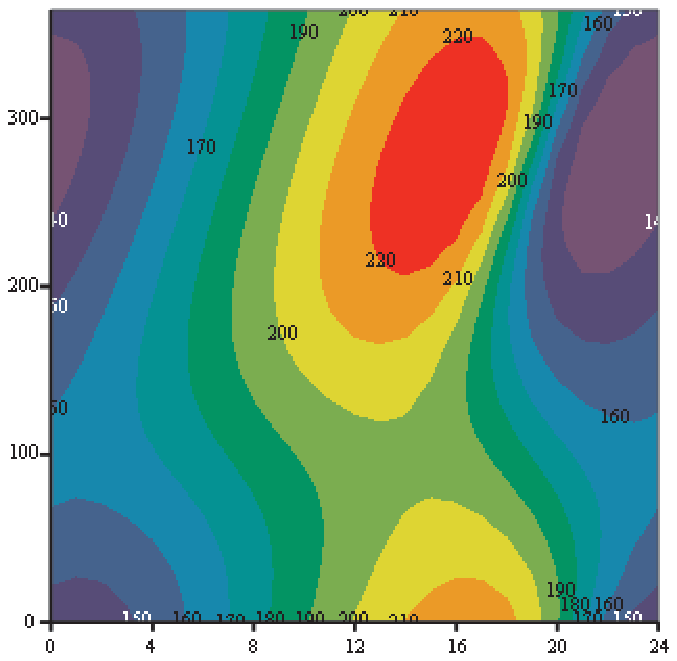}}}
\end{picture}
\caption{\label{FigEffectMiller2} 
Ether effect for an observer at a latitude of 51.8$^o$ and a longitude of 4.6$^o$ and a Sun velocity according to Miller 2 in table~\ref{tabSun} as function of sidereal time (horizontal scale) and days after 1 January 2015 (vertical scale). Left: In-plane-of-interferometer ether velocity component (in 0.001 c); Right: Azimuth of maximum of effect upon rotation (in degree).
}
\end{figure} 

For an observer at a latitude of 51.8$^o$ and a longitude of 4.6$^o$ (i.e. location Puttershoek in the Netherlands) and a Sun velocity according to Miller 2 in table~\ref{tabSun} the magnitude of the Ether velocity component in the plane of rotation of a horizontal interferometer ($\hat{\beta}$) and the Azimuth at which this maximum occurs are shown in figure~\ref{FigEffectMiller2}. It is clear that the effect changes depending on the sidereal time (horizontal scale) and season (vertical scale) as represented by the day after January 1, 2015. Both the magnitude of the maximum and the direction in which the maximum occurs change.

\section{Experiment}

\begin{figure}
\begin{picture}(450,300)
\put(75,0){\scalebox{1}{\includegraphics{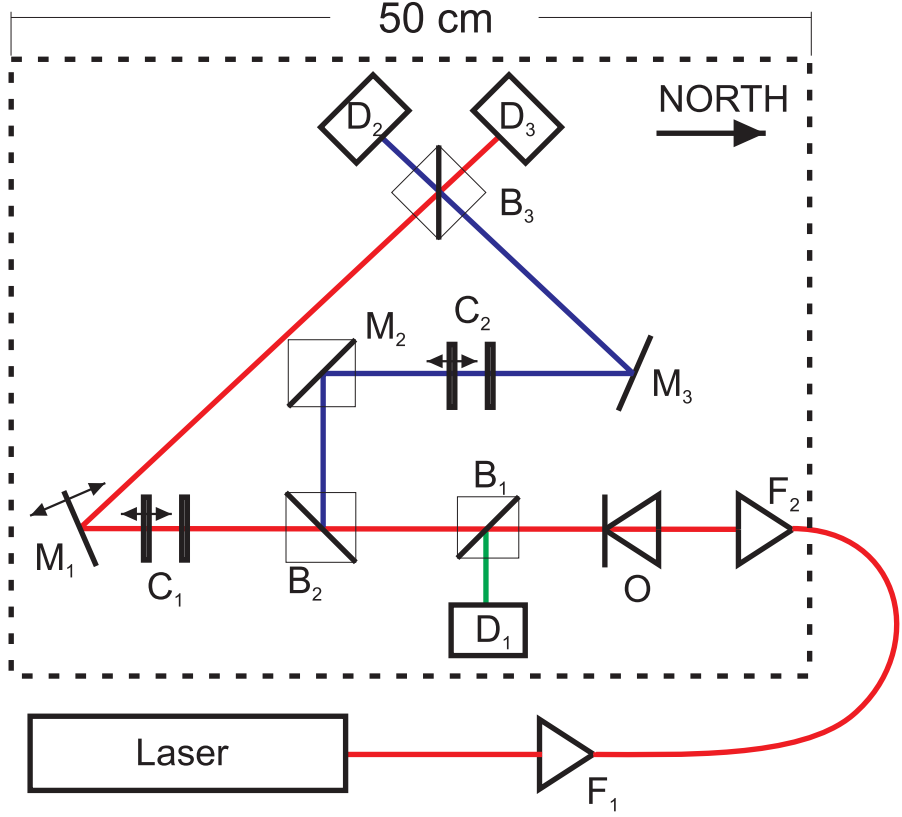}}}
\end{picture}
\caption{\label{ExpSetup} 
Experimental set-up. Locations to scale.}
\end{figure}

The experimental set-up used is shown in figure~\ref{ExpSetup}. 
It is based on the assumption that the effect is first-order in the velocity with respect to the ether. When the light beams travel in opposite direction also the effect is opposite. When the Mach-Zehnder interferometer is constructed as shown in the figure then the effect doubles. The light of a laser (a stabilized He-Ne laser type Coherent 200, linear polarized, 0.5~mW, maximum mode sweep of 10~MHz) is coupled into a fiber optical cable (Patch Cable, FC/PC, 630~nm, PM Panda Style, 1~m, Thorlabs) by means of fiberport $F_1$ (FiberPort PAF-X-5-B, FC/PC \& APC, EFL=4.6~mm, 600-1050~nm, Thorlabs). The fiber optical cable is led into a temperature stabilized containment schematical denoted by the dashed box. The temperature of this box is stable for days within 3~mK. For shorter times the temperature is stable within 1~mK. The heat applied to control the temperature of the box varies between 1 and 4~W. This ensures that the possible temperature gradients within the inner compartment are less than 1~mK/m. Inside the box the laser light exits another fiberport $F_2$ (similar to $F_1$). To prevent optical feedback from the interferometer into the laser a polarization dependent optical isolator $O$ (IO-2D-633-VLP from Thorlabs, isolation at least 35~dB) was included. After transmission through the optical isolator the light beam is split into two arms at beam splitter $B_1$ (Cube Mounted Polarization Insensitive Beam splitter, 400-700~nm, CM1BS013 from Thorlabs). The first arm (green line in the figure) is directed to a light detector $D_1$ (all detectors used are amplified Silicon detectors, PDA36A from Thorlabs). This detector is used to monitor the output intensity of the laser. The other arm continues to a second beam splitter $B_2$ where the light beam is split in arms A (red line) and B (blue line) of the Mach-Zehnder interferometer. These light beams are joined together at the third beam splitter $B_3$ after which the light beam interference is detected by detectors $D_2$ and $D_3$ as described in section~\ref{secMZ}. The light in arm A passes a Fabry-P\'{e}rot cavity $C_1$ similar to the one described in section~\ref{secFB}, reflects at mirror $M_1$ (12.7~mm diameter Protected Silver Mirror, 6.0~mm thick, Thorlabs PF0503P01) which is connected to a piezo crystal (Piezoelectric Actuator, Max Displacement 9.1~$\mu$m, 3.5 x 4.5 x 10 mm, AE0203D08F Thorlabs) to be able to adjust the phase difference of the light at beam splitter $B_3$. The light in arm B reflects at mirror $M_2$ (Cube Mounted Protected Silver Turning Mirror, Thorlabs CM1P01) passes a Fabry-P\'{e}rot cavity $C_2$ similar as before, reflects at mirror $M_3$ (similar to $M_1$) into the direction of beam splitter $B_3$ where it is joined to the light of arm A. One window of each Fabry-P\'{e}rot cavity is mounted by means of a piezo crystal (as before). All mirrors and windows are mounted with kinematic mounts to enable angle adjustments to make sure that the beams interfere in the cavities and at beam splitter $B_3$. The set-up is designed in such a way that the lengths of arm A and B are approximately equal (both 464~mm within a few~mm). This limits the influence of laser instabilities on the measured phase difference. The phase is measured with an accuracy of approximately 0.03~rad. The influence of the temperature of the box on the phase was measured by means of an oscillating temperature (with an amplitude of 0.04~K and a period of 2.7 hours) and was 65(1)~rad/K. This explains the need for a stable temperature. The transmission of both cavities is kept minimal by adjustments of the cavity length by means of the piezo crystals. In this way the transmission is kept minimal within 0.3~\% of the transmission.

\begin{figure}
\begin{picture}(450,600)
\put(2,0){\scalebox{0.6}{\includegraphics{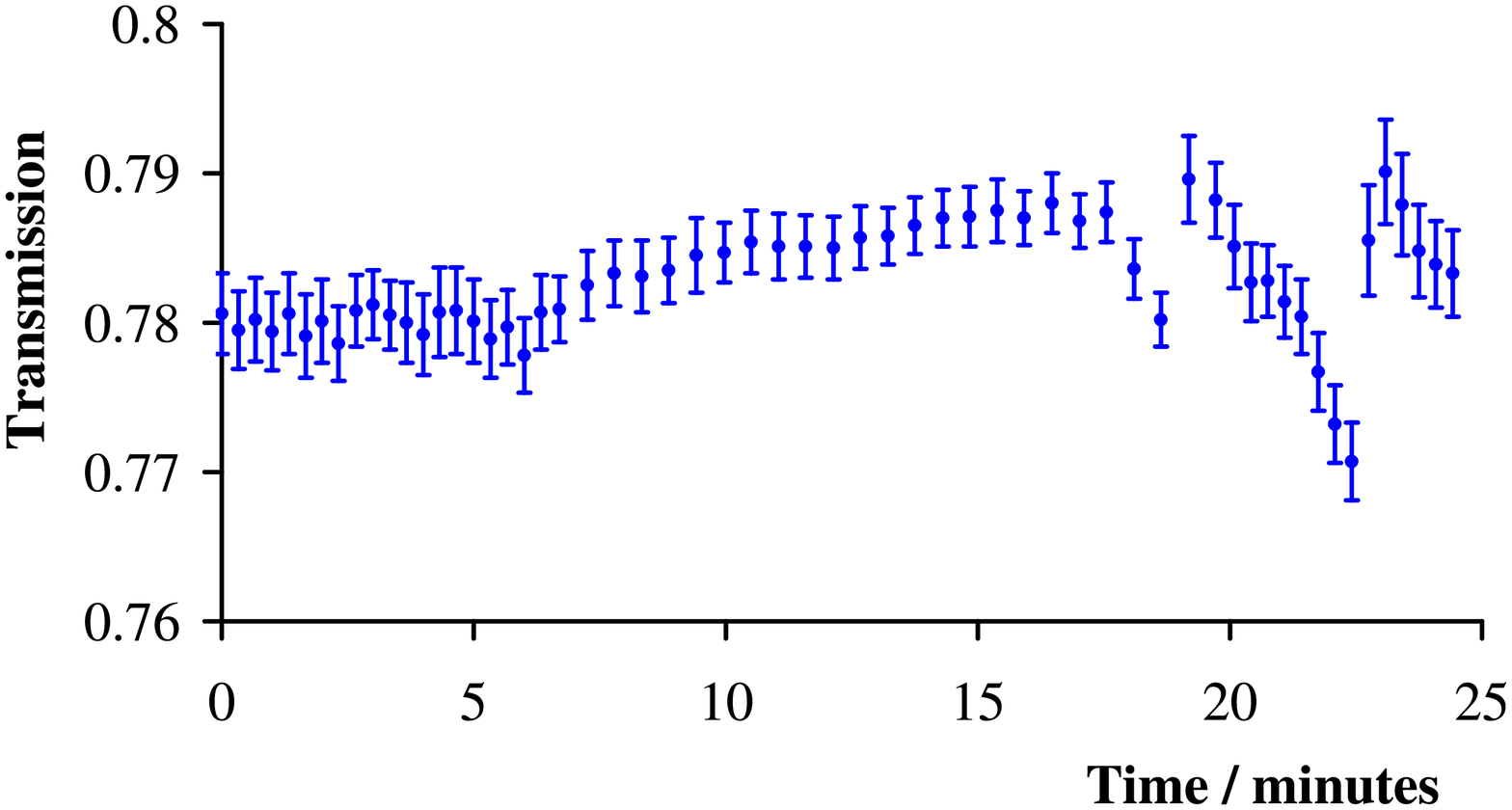}}}
\put(9,210){\scalebox{0.584}{\includegraphics{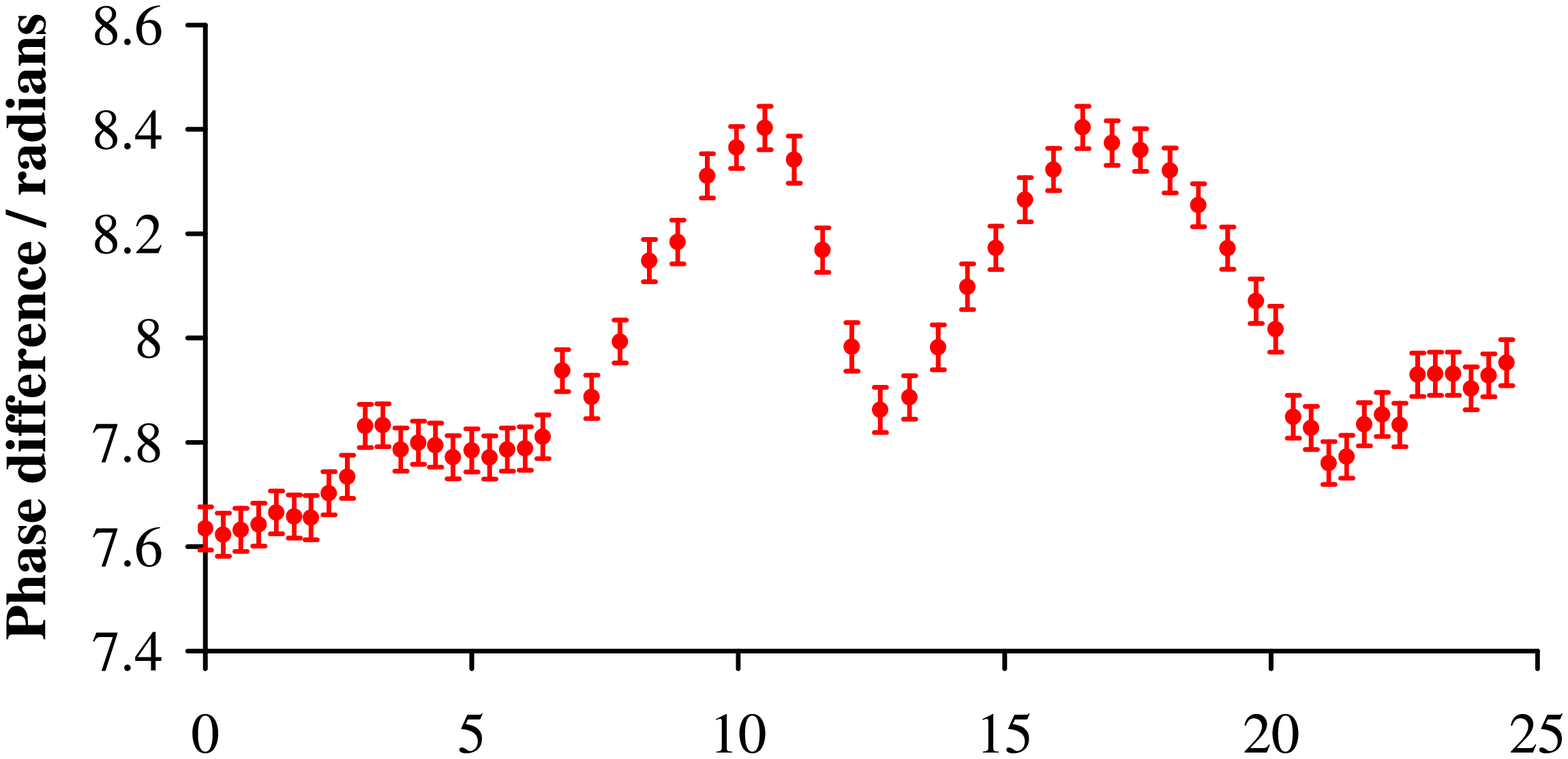}}}
\put(9,400){\scalebox{0.6}{\includegraphics{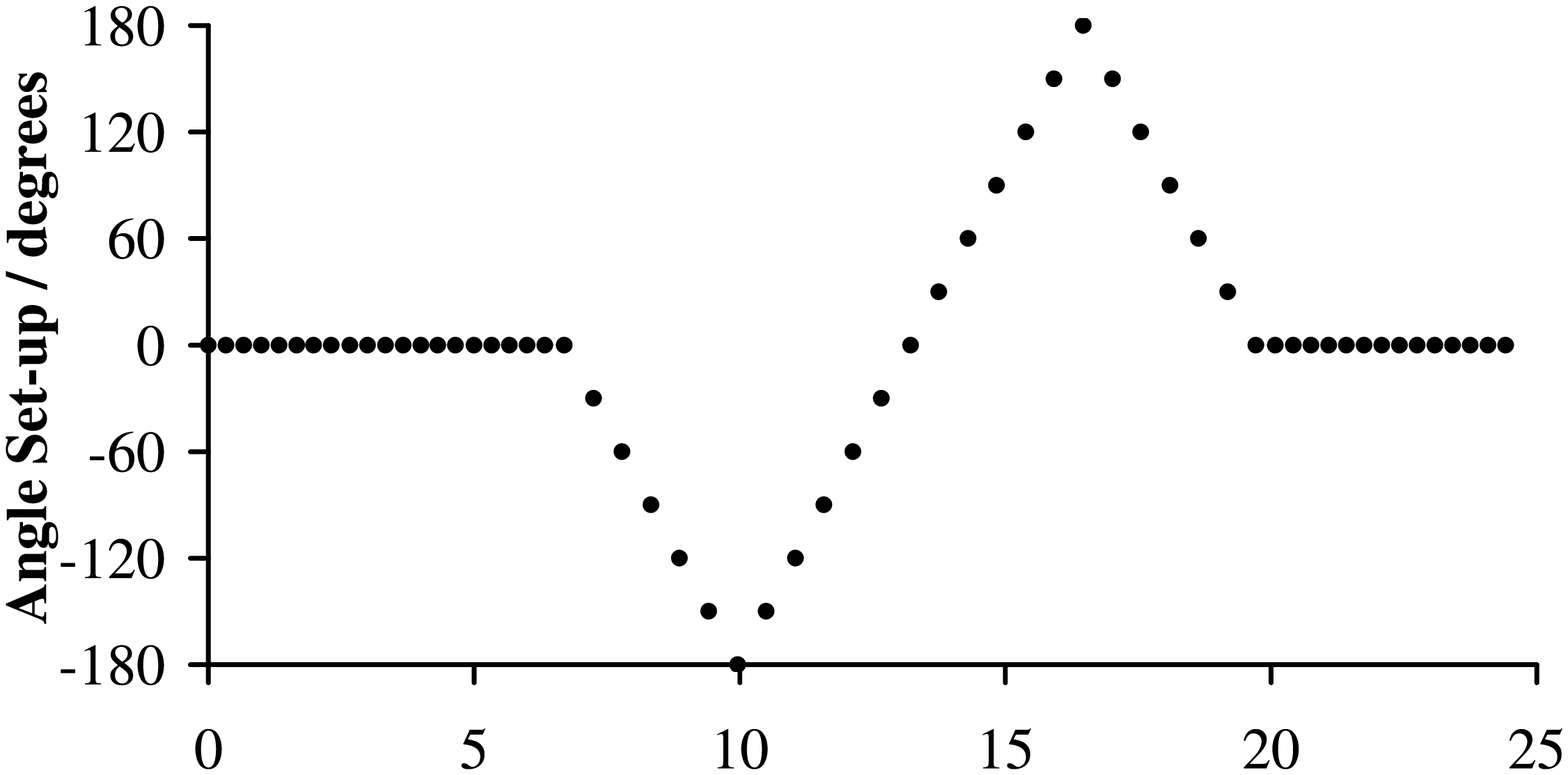}}}
\end{picture}
\caption{\label{figRotationExample} Top: angle of set-up as function of time. Middle: phase difference of interferometer during the same period as the upper graph. Bottom: Transmission of the interferometer again during the same period.}
\end{figure}

For determination of the effect, the set-up is rotated along a vertical axis. Every hour the set-up rotates from 0 to -180 degrees, from -180 to +180 and from +180 to 0 with steps of 30 degrees. For 0 degrees the interferometer is directed according to figure~\ref{ExpSetup}. For 90 degrees the arrow in this figure points to the local West. At every step the phase difference and transmission of the interferometer are measured during 20 seconds to average out statistical variations. A complete rotation takes about 20 minutes. A typical response of the interferometer as function of angle is shown in figure~\ref{figRotationExample}.
The top graph of this figure shows the angle of the set-up as function of time. The middle one shows the phase difference of the interferometer during the same period as the upper graph. The bottom graph shows the transmission of the interferometer again during the same period. In the lower graph the phase oscillations due to the rotation of the set-up are clearly visible. 

\section{Results}
At first, it seems that this interferometer gives an indication that there might be some effect on the phase difference due to the motion of the Earth, but the transmission is completely independent of the angle of the set-up. This corresponds to the anticipated effect. However, if such an effect exists it should depend on the time of day and year. Upon rotation of the set up, the amplitude and azimuth of the maximum should vary between certain minima and maxima depending on the orientation of the Earth velocity with respect to the preferred frame as discussed in the previous section. The Fourier transform of the phase difference (corrected for the linear assumed drift) as function of rotation angle gives this amplitude and azimuth of the maximum phase difference for all orders. The zeroth order is just the average phase difference during a rotation. The first order represents the amplitude and azimuth of that part of the signal that varies with the cosine of the angle of the set-up, corresponding to first order effects in $v/c$. The second order represents the amplitude and azimuth of that part of the signal that varies with the cosine of twice the angle of the set-up, corresponding to second order effects, and so on. The error in the values can be estimated from the difference between the Fourier transform of the data points measured for increasing set-up angles and the one measured for decreasing set-up angles (to find the systematic error due to the unknown drift) combined with the Fourier transform of the variances (to find an estimate of the statistical error). The amplitude and azimuth should vary with the sidereal time and epoch as discussed in the previous section. 

\begin{figure}
\begin{picture}(400,600)
\put(2,0){\scalebox{0.8}{\includegraphics{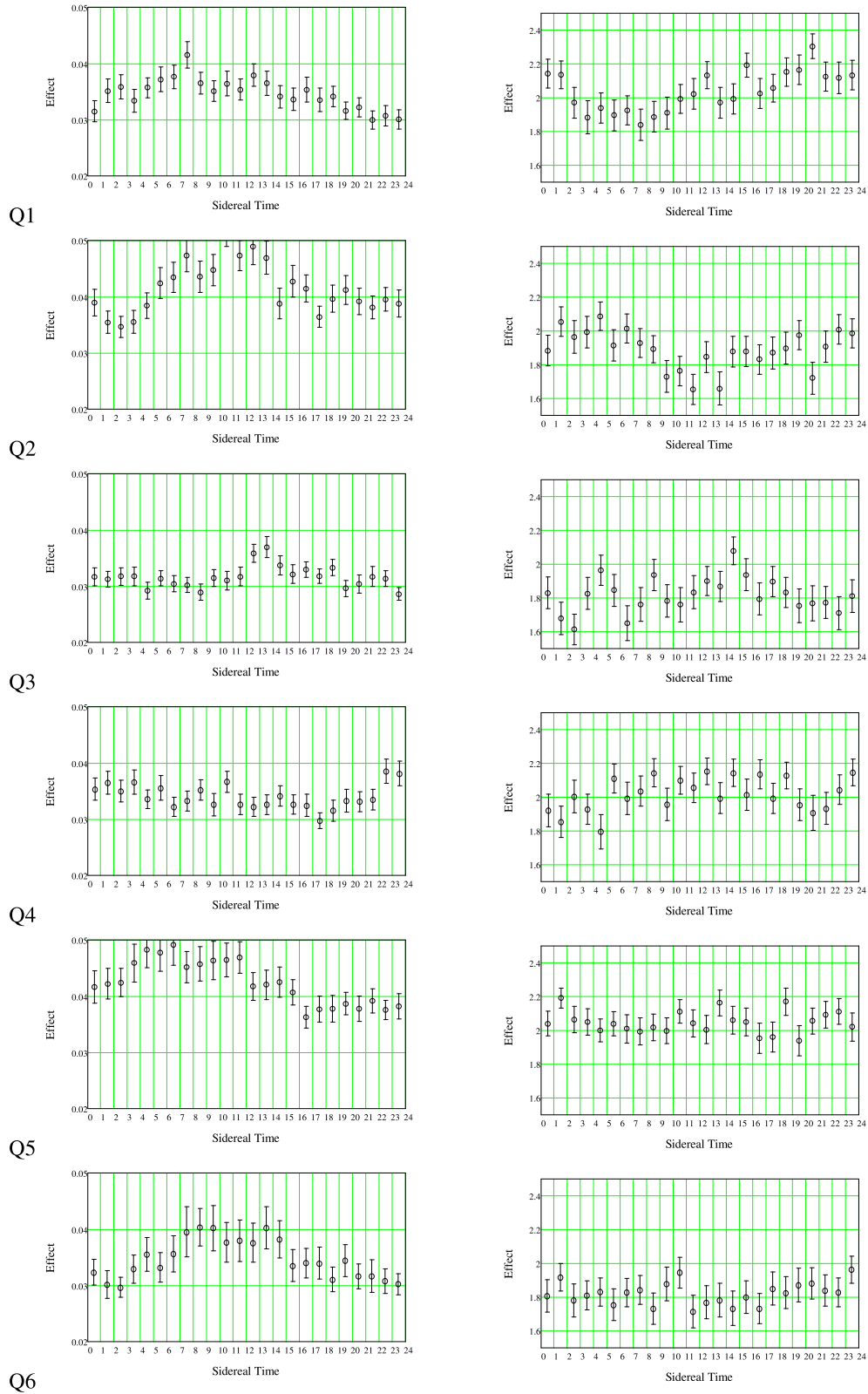}}}
\end{picture}
\caption{\label{FigData} Amplitude of phase difference upon rotation (left) and angle of maximum phase difference upon rotation (right) as function of sidereal time for 6 consecutive quarter years from April 2013 until November 2014.}
\end{figure}

The results for 6 consecutive quarter years (Q1 to Q6) are shown in figure~\ref{FigData}. Although some variation is observed, the quality of the data is not sufficient to reach a final conclusion. The variations are less then are expected. This can have several causes as the absence of the effect or a masking effect or an wrong expectation of the magnitude of the effect.

It is possible to increase the sensitivity of the measurements by increasing the stability of the set-up (for instance mechanical stability and laser stability), reduce the rotation time of the set-up to reduce environmental effects and so on. It is recommended that the measurements are repeated at a higher altitude (as Miller has done) to reduce the influence of the possible entrainment of the ether that can reduce the effect considerably (note that the height above see level of Puttershoek is -10 m).

\section{Conclusion}
A novel experiment has been performed to observe possible deviations from the predictions of Einstein's special relativity theory. Although the experiment reveals some sidereal deviations, the magnitude of the measured deviations is too small to reach any final conclusion. It is recommended that the experiment is repeated with a more stable set-up at higher altitudes.
\bibliography{main}

\end{document}